\documentclass[sigconf]{acmart}

\AtBeginDocument{%
  \providecommand\BibTeX{{%
    \normalfont B\kern-0.5em{\scshape i\kern-0.25em b}\kern-0.8em\TeX}}}

\copyrightyear{2024}
\acmYear{2024}
\setcopyright{acmlicensed}\acmConference[ASSETS '24]{The 26th International ACM SIGACCESS Conference on Computers and Accessibility}{October 27--30, 2024}{St. John's, NL, Canada}
\acmBooktitle{The 26th International ACM SIGACCESS Conference on Computers and Accessibility (ASSETS '24), October 27--30, 2024, St. John's, NL, Canada}
\acmDOI{10.1145/3663548.3675620}
\acmISBN{979-8-4007-0677-6/24/10}

\usepackage{subfigure}

\usepackage{tabularx}
\usepackage{array}
\usepackage{ragged2e}
\newcolumntype{P}[1]{>{\RaggedRight\hspace{0pt}}p{#1}}

\begin{document}

\raggedbottom

%%
%% The "title" command has an optional parameter,
%% allowing the author to define a "short title" to be used in page headers.
\title[“I Try to Represent Myself as I Am”: Self-Presentation Preferences of \\People with Invisible Disabilities through Embodied Social VR Avatars]{“I Try to Represent Myself as I Am”: Self-Presentation\\ Preferences of People with Invisible Disabilities through\\ Embodied Social VR Avatars}

\author{Ria J. Gualano}
\authornote{Both authors contributed equally to the work.}
\affiliation{%
  \institution{Cornell University}
  \city{Ithaca, NY}
  \country{USA}}
\email{rjg322@cornell.edu}

\author{Lucy Jiang}
\authornotemark[1]
\affiliation{%
  \institution{Cornell University}
  \city{Ithaca, NY}
  \country{USA}}
\email{lucjia@cs.cornell.edu}

\author{Kexin Zhang}
\affiliation{%
  \institution{University of Wisconsin-Madison}
  \city{Madison, WI}
  \country{USA}}
\email{kzhang284@wisc.edu}

\author{Tanisha Shende}
\affiliation{%
  \institution{Oberlin College}
  \city{Oberlin, OH}
  \country{USA}}
\email{tshende@oberlin.edu}

\author{Andrea Stevenson Won}
\affiliation{%
  \institution{Cornell University}
  \city{Ithaca, NY}
  \country{USA}}
\email{asw248@cornell.edu}

\author{Shiri Azenkot}
\affiliation{%
  \institution{Cornell Tech}
  \city{New York, NY}
  \country{USA}}
\email{shiri.azenkot@cornell.edu}

\renewcommand{\shortauthors}{Gualano and Jiang et al.}
%%
%% The abstract is a short summary of the work to be presented in the
%% article.
\begin{abstract}
With the increasing adoption of social virtual reality (VR), it is critical to design inclusive avatars. While researchers have investigated how and why blind and d/Deaf people wish to disclose their disabilities in VR, little is known about the preferences of many others with invisible disabilities (e.g., ADHD, dyslexia, chronic conditions). We filled this gap by interviewing 15 participants, each with one to three invisible disabilities, who represented 22 different invisible disabilities in total. We found that invisibly disabled people approached avatar-based disclosure through contextualized considerations informed by their prior experiences. For example, some wished to use VR’s embodied affordances, such as facial expressions and body language, to dynamically represent their energy level or willingness to engage with others, while others preferred not to disclose their disability identity in any context. We define a binary framework for embodied invisible disability expression (public and private) and discuss three disclosure patterns (Activists, Non-Disclosers, and Situational Disclosers) to inform the design of future inclusive VR experiences.

\end{abstract}

%%
%% The code below is generated by the tool at http://dl.acm.org/ccs.cfm.
%% Please copy and paste the code instead of the example below.
%%


%%
%% Keywords. The author(s) should pick words that accurately describe
%% the work being presented. Separate the keywords with commas.

%\received{20 February 2007}
%\received[revised]{12 March 2009}
%\received[accepted]{5 June 2009}

%%
%% This command processes the author and affiliation and title
%% information and builds the first part of the formatted document.
\maketitle
\section{Introduction}
Like anyone else, disabled people may want to represent their unique identities through their avatars when interacting with others in social virtual reality (VR) spaces. Social VR platforms offer users a strong sense of embodiment and freedom from some forms of physical constraint. As more people turn to VR for social connection \cite{maloney2020falling, yalon2008virtual}, the importance of diverse mechanisms for self-expression grows. However, the accessibility research community has yet to fully understand how disabled social VR users wish to represent various disability identities through embodied avatars. 

Among other functions, avatars are vessels for self-expression in VR, and their characteristics provide cues that mediate interpersonal interactions \cite{freeman2021body, won2023your}. Previous research has investigated the avatar representation preferences of people with visible disabilities (i.e., disabilities that are apparent due to the presence of physical characteristics or assistive technologies \cite{cdcdefinition}) on social VR platforms. For example, Zhang et al. \cite{zhang2022s} found that d/Deaf and hard of hearing people and blind and low vision people wanted to disclose their disabilities and raise awareness through adding assistive technology to their avatars. 

However, many disabled people do not have characteristics that are immediately visible \cite{invisibledisabilitydef}. For the 33 million people in the United States alone who have invisible disabilities (approximately 10\% of the U.S. population) \cite{disabledworld}, they may not use assistive technologies, or their assistive technologies may not be recognizable or apparent to others. As such, prior findings about disability representation through social VR avatars do not readily apply to their experiences. 

More recently, Mack et al. \cite{mack2023towards} studied how people with both visible and invisible disabilities prefer to represent themselves on social media platforms (e.g., Snapchat, Facebook). The authors found that visibly disabled users wanted their avatars to show their full bodies and assistive technologies, while invisibly disabled users preferred to use symbolism (e.g., words in the dyslexia-friendly font “Bionic”) to signify their disabilities. However, they only briefly explored invisible disability avatar representation on digital platforms. At present, little is known about invisibly disabled users’ unique perspectives on when, why, and how to disclose through embodied avatars in social VR spaces. 

To address this gap, we investigated people with invisible disabilities’ current representation practices and desired representations, as well as how these preferences might differ across social VR contexts. Specifically, we posed the following research questions: 

\begin{itemize}
	\item \textbf{RQ1:} How do people with invisible disabilities feel about representing themselves through embodied vs. non-embodied avatars?
	\item \textbf{RQ2:} How do invisibly disabled people represent their disability through social VR avatars today, and how do they wish to represent themselves through social VR avatars in the future?
	\item \textbf{RQ3:} In what social VR contexts (when and why) do invisibly disabled people wish to disclose their disability identities and experiences?
\end{itemize}

We conducted semi-structured interviews with 15 participants with invisible disabilities such as ADHD, dyslexia, and chronic conditions. We asked them about their current avatars, the similarities and differences between their embodied avatars and their non-virtual appearance, and the different contexts in which they engaged in social VR. During the interviews, we included an interactive brainstorming component where participants drew, typed, or otherwise visualized their \textit{ideal} avatar self-representation, going beyond what is currently available on customization platforms. 

We found that social VR’s embodied nature impacted invisibly disabled users’ disclosure preferences and methods. We identified two distinct methods through which invisibly disabled people represented their identity: \textbf{public} (i.e., explicit representations like disability-related shirt graphics) and \textbf{private} (i.e., implicit representations like knee pads that symbolize a need for self-protection from bipolar mood swings). Through participants’ disclosure patterns, we also identified three groups of disclosers: \textbf{Activists}, \textbf{Non-Disclosers}, and \textbf{Situational Disclosers}. While some perceived social VR to be a safer space than non-virtual social spaces and felt comfortable disclosing through customizable accessories, others preferred not to disclose due to their heightened discomfort associated with harassment within embodied experiences. Based on our findings, we conclude with design recommendations to guide more inclusive avatar representation.

This paper is an extension of our work-in-progress poster published at ASSETS 2023 \cite{gualano2023invisible}. We contribute novel insights into how avatar customization can support the large and underserved population of people with invisible disabilities. Through this work, we amplify the experiences of this historically underrepresented group within a social VR context, taking another step toward making social settings more inclusive to all users. 

\section{Related Work}
Disability representation is important for people with both visible and invisible disabilities. Broadly, social VR platforms lack diverse customization options (e.g., body and racial diversity \cite{lee2014does, lee2011whose, morris2023don}). Invisible disability representation in social VR is even less present. Social VR offers unique affordances for users to represent aspects of their identities in different ways than they can in non-virtual contexts \cite{freeman2021body}. By allowing users to render and share their typically invisible experiences of disability, VR also provides new advocacy opportunities. We build on prior literature regarding contextual disability disclosure, social VR experiences, and avatar-based identity expression. To our knowledge, we are the first to explore the nuances of disability disclosure across a wide range of invisible disabilities in embodied social environments. 

\subsection{Disability Disclosure in Employment, Academic, and Social Settings}
Although discrimination against disabled people is legally prohibited\footnote{Both the Americans with Disabilities Act (ADA) of 1990 and Section 504 of the Rehabilitation Act of 1973 protect people from discrimination on the basis of a disability.}, discrimination and ableism are still prevalent in today’s society. As a result, disabled people often carefully consider whether, when, and how to disclose their disabilities. 

At school and in the workplace, disabled people may require accommodations to perform their roles successfully; however, disclosure also heightens the risk of experiencing disability-related discrimination \cite{lindsay2018systematic, lindsay2019disability, wilton2006disability, von2014perspectives, dalgin2008invisible}. In addition to fears of potential social rejection by their peers and supervisors \cite{lindsay2019disability, lindsay2018systematic}, disabled people may also worry that disclosure will indirectly impact their academic and professional success \cite{rocco2004towards, toth2022disclosure, toth2014employee, mcgrath2023disclosure, gignac2022workplace, pearson2003to}. Prior literature has found that people with invisible disabilities can conceal their disabilities more easily than people with visible disabilities, especially if they perceive their conditions as “mild” or “irrelevant” to their performance \cite{lindsay2019disability, von2014perspectives, dalgin2008invisible}. 

While there is substantial work investigating disabled peoples’ disclosure strategies in employment and academic environments, there is comparatively less research on disclosure in technology-mediated social settings. Some researchers have explored the nuances of disability disclosure specifically for online dating \cite{porter2017filtered, saltes2013disability}, while others have focused on social media more broadly \cite{furr2016strategic, kaur2022social, heung2024vulnerable}. For example, Furr et al. \cite{furr2016strategic} explored visibly disabled people’s disclosure strategies on social media. They found that the curatorial nature of social media meant participants’ visible disabilities were not immediately apparent unless disclosed through text or images. Based on these insights, the authors identified three approaches: open (full disability disclosure to a broad audience), secure (“routine” disability disclosure to a private audience through posts or status updates), and limited (selective disclosure across all audiences). Participants who adopted an open disclosure approach likened disclosing their disability to “coming out” on social media, whereas those who were more hesitant avoided disclosure due to prior ableist experiences. Similarly, in a study of disabled content creators’ practices on social media, Heung et al. \cite{heung2024vulnerable} found that some content creators strategically censored or limited their online presence to mitigate disability-related hate and harassment.

Distinct from the experiences of people with visible disabilities, the non-apparent nature of invisible disabilities means that invisibly disabled people may not be taken as seriously when seeking accommodations or outwardly expressing their disability. Evans et al. \cite{evans2019trial} found that a single individual often employed multiple disclosure strategies, dependent on context, and disclosure decisions were often guided by a need to assert access needs in physical environments (e.g., reading small text, avoiding severe food allergens). Additionally, invisibly disabled people may anticipate negative perceptions stemming from a lack of understanding of their disabilities, which may make them hesitant to disclose \cite{mullins2013lived, morina2022when}. While people with both visible and invisible disabilities anticipate and encounter stigma, invisibly disabled people may struggle more with validating their disability to themselves and others. This can be especially consequential given the fluctuating (in)visibility of disability; if assistive technologies become apparent in social interactions, they can render invisible disabilities visible \cite{faucett2017visibility}. 

These studies provide necessary groundwork for understanding disability disclosure in emerging social settings, such as social media. Our empirical study provides additional insight into the experiences and challenges invisibly disabled people face when deciding to represent or otherwise disclose their disability in the virtual world.

\subsection{Social Virtual Reality}
Social VR, which enables immersive 3D experiences by allowing users to meet, interact, and socialize in the form of virtual avatars \cite{freeman2021body}, has recently gained popularity. Unlike 2D social media and virtual worlds, social VR engages users in embodied experiences through head-mounted displays, which allow real-time body movements and gestures to be tracked and rendered in virtual environments \cite{caserman2019real}. In virtual reality environments, a user’s sense of embodiment is characterized by their ability to control their body’s movements, experience sensations as their virtual body, and feel as though they are within their body \cite{kilteni2012sense}. VR can facilitate a greater sense of embodiment and social presence than other non-VR platforms \cite{maloney2020falling, tham2018understanding}. 

With these unique affordances, social VR platforms offer a variety of avatar-mediated activities and social interactions. To understand how people engage in social VR, Maloney and Freeman \cite{maloney2020falling} interviewed 30 social VR users, finding that they enjoyed experiencing everyday activities in immersive ways. Users connected with others by hosting and attending gatherings, playing games together, and working remotely \cite{freeman2021body, abramczuk2023meet}. 

Social VR can be appealing to disabled people since they can address access needs in ways that are not possible in the non-virtual world. For example, VR enables people with physical and intellectual disabilities to engage in active leisure activities, such as snowboarding, which may be inaccessible to them in the physical world \cite{yalon2008virtual}. VR can also make social settings more accessible. Prior work has found that neurodivergent users have greater control over their environment in VR and may remove themselves from overwhelming situations or adjust settings to reduce their sensory load \cite{boyd2018leveling}. Given social VR’s increasing popularity and novel affordances, it is important to understand how invisibly disabled people wish to represent themselves within this space.

\subsection{Traditional and Creative Avatar Expression}
Avatars enable users to develop a greater social presence in social VR. Prior work has found that avatar realism, through both appearance and behavior, can improve a user’s sense of immersion and interpersonal understanding \cite{roth2016avatar, steed2015collaboration, won2023your}. Furthermore, Langener et al. \cite{langener2022immersive} determined that avatar customization heightens a user’s sense of connection and ownership over their virtual body. In studying selective self-presentation in social VR, Freeman et al. \cite{freeman2021body} identified that many social VR users strove to make their avatars as similar as possible to their non-virtual counterparts because they considered avatars an extension of themselves in virtual worlds. Won et al. \cite{won2023your} established a design space for understanding avatar representation in different social contexts and for different purposes, ranging from low anonymity and low immersion in professional contexts to high anonymity and high immersion for entertainment. 

Despite some progress, current avatars are generally inadequate at capturing the wide range of human appearances and identities. Marginalized and stigmatized groups are especially under-represented. For example, Ready Player Me \cite{readyplayerme}, a popular avatar creation tool, presents gender as a binary, which neglects the diversity of gender identity. Researchers have explored identity and avatar representation preferences in social VR, including diverse user groups such as racial minorities \cite{blackwell2019harassment}, women \cite{sharvit2021virtual}, and LGBTQ users \cite{freeman2022acting, freeman2022re, povinelli2024springboard}. Some have specifically explored the avatar choices of non-cisgender users, revealing that they changed their avatar’s gender expression frequently through outfits and accessories to indicate their flexible and gender-fluid identities \cite{freeman2022re} and used voice changers to express their gender identities in social VR \cite{povinelli2024springboard}. 

In addition to representing a person’s appearance, avatars are often used to creatively express users’ interests and identities, sometimes extending beyond what is possible in the physical world. Users may opt for fantastical, cartoonish, and non-human features, such as the red or snake-like eyes offered in Ready Player Me \cite{readyplayerme}. Alternatively, they can choose non-humanoid avatars altogether. Furthermore, users can enable complete or partial invisibility of their avatars by manipulating the meshes that make up the 3D model \cite{youtube_invisibility_toolbox}. When exploring the contextual dynamics influencing users’ avatar choices, Ribeiro et al. \cite{ribeiro2024towards} found that people most often used avatars that closely resembled their actual appearance in work or academic meetings, but preferred to have a different representation when interacting or gaming with strangers. With the flexibility of presenting as they are or as they would want to be, users can communicate their uniqueness, explore life as their avatar, and maintain greater control over how others perceive them. 

Avatar representation of disability is particularly understudied, with some exceptions. Zhang et al. \cite{zhang2022s} found that users wished to disclose disabilities by attaching assistive technology to their avatars (e.g., hearing aids) and generally wished to align their avatar representations with their non-virtual appearances. Additionally, Davis and Chansiri \cite{davis2019digital} found that online users with disabilities customized their avatars to foster a greater sense of empowerment and embodiment. For instance, one user depicted herself as younger because she felt that her disability removed years from her life. Regarding avatar presentation across digital social media platforms, Mack et al. \cite{mack2023towards} interviewed individuals with intersectional identities, finding that they often had to choose between representing their disability or race-related identities in avatars. They also found that disabled people wanted to use symbolic representations that would resonate with their specific community (e.g., chronic conditions, ADHD). 

Our study includes social VR users with invisible disabilities, distinct from Zhang et al.’s study \cite{zhang2022s} which focused on people with visible disabilities. Though many participants with visible disabilities were interested in translating their non-virtual assistive technology into their virtual avatars, little is known about the preferences of invisibly disabled people. In addition, we build on Mack et al.’s study \cite{mack2023towards} on the social media avatar preferences of participants with visible and invisible disabilities. We focus our inquiry on invisible disability representation preferences in multimodal and embodied virtual contexts, filling a gap in existing literature. 

By investigating the preferences of people with invisible disabilities, we shift the focus from translating assistive technology into the virtual world and instead explore social VR affordances for illuminating experiences that are not necessarily visible in non-virtual spaces. In this study, we examine how users want to employ VR’s multimodal affordances for disability expression in spaces where avatars mediate all social interaction. 

\section{Methodology}
We conducted and analyzed data from semi-structured interviews with 15 participants who self-identified as having one or more invisible disabilities. We aimed to understand their perspectives on disability representation and disclosure in social VR settings.

\subsection{Participants}
We recruited 16 participants by emailing disability-related organization mailing lists, posting on Reddit, and partnering with our university’s undergraduate extra credit system. We recruited until we reached saturation, a common practice when conducting thematic analysis \cite{braun2006using}. Our inclusion criteria indicated that participants had to be 18 years old or older and have an invisible disability, defined as a physical or psychological condition with no visible manifestation. We welcomed participants who self-identified as invisibly disabled (no diagnosis required). Participants were also required to have prior experience using social VR platforms such as Rec Room, VRChat, vTime XR, and AltspaceVR (note: recruitment began in January 2023, before AltspaceVR was shut down). We defined prior experience as having used social VR frequently enough to be familiar with their current avatar customization processes. Due to inconsistent identification (i.e., different names during and after our study), we discarded data from one participant. 

The remaining 15 participants’ ages ranged from 20 to 56 years old ($mean = 33.07, SD = 9.76$). Eight identified as female and seven identified as male. Regarding race, seven participants identified as white, four as Black or African American, two as mixed-race, one as Asian, and one as Alaskan Native. All participants disclosed that they had at least one invisible disability, such as dyslexia, chronic conditions, ADHD, or autism. Some participants had multiple disabilities; for example, P15 had ADHD, anxiety, and depression. Two participants did not disclose their specific disabilities but did share the general type of disability (e.g., chronic condition, learning disability). In total, 22 unique invisible disabilities were represented. All participants had used at least one social VR platform, but many used multiple. Most used VR on a daily or weekly basis. We present a detailed demographic breakdown in Table \ref{table:participants}. 

Participants were compensated with a \$20 digital gift card upon completion of the interview. This study and all procedures were approved by our university’s Institutional Review Board (IRB).

\begin{table*}[ht!]
\renewcommand{\arraystretch}{1.15}
\begin{center}
\caption{Participant demographics, including their self-reported age, gender, race, and invisible disability identity. We also present the social VR platforms they use and frequency of VR use.}
\label{table:participants}
\begin{tabular}{>{\raggedright}p{0.03\textwidth} >{\raggedright}p{0.11\textwidth} >{\raggedright}p{0.09\textwidth} >{\raggedright}p{0.22\textwidth} >{\raggedright}p{0.23\textwidth} >{\raggedright\arraybackslash}p{0.2\textwidth}}
\toprule
\textbf{ID} & \textbf{Age / Gender} & \textbf{Race} & \textbf{Invisible Disabilities} & \textbf{Social VR Platforms} & \textbf{Frequency of Usage} \\
\midrule[\heavyrulewidth]
\textbf{P1} & 32 / Female & White & Low vision; chronic condition & AltspaceVR; Spatial & Several times a week \\
\textbf{P2} & 32 / Female & Alaskan native & Bipolar disorder; PTSD & Synth Riders; Fishing; VRChat; Mini Golf  & Daily \\
\textbf{P3} & 56 / Female & Mixed-race & Dyslexia; rheumatoid arthritis; congestive heart failure & AltspaceVR; VR fishing; VRChat; Spatial; Big Screen; Wander & Weekly \\
\textbf{P4} & 38 / Female & White & Dyslexia; epilepsy; chronic pain & FitXR, Synth Riders, Beat Saber, VRChat, AltSpaceVR, BanterVR & Two to four or five times a week \\
\textbf{P5} & 31 / Male & White & RSI; ADHD; Asperger’s & VRChat; Rec Room & Social VR: monthly, VR in general: weekly \\
\textbf{P6} & 24 / Female & Black & POTS & vTime XR; AltspaceVR; Rec Room; Ping Pong & Weekly \\
\textbf{P7} & 33 / Female & White & Depression & VRChat & Daily \\
\textbf{P8} & 20 / Male & Asian & Chronic condition (diagnosis undisclosed) & VRChat & One to two times a month \\
\textbf{P9} & 28 / Male & White / Asian & Autism (nonverbal); learning impairment with linguistics (text); bipolar type one & VRChat & “Extensive”; 5000-6000 hours total \\
\textbf{P10} & 52 / Female & White & Hashimoto’s disease; contextual social anxiety; osteoporosis & VRChat, Synth Riders & “Live on there” - work, sleep, “addiction” \\
\textbf{P11} & 28 / Male & Black & ADHD (undiagnosed) & AltspaceVR; Spatial; Rec Room; Horizon Worlds & Multiple times a month \\
\textbf{P12} & 31 / Male & White & Asperger’s; ADHD & AltspaceVR; Oculus; Rec Room; Facebook Horizon & Uses as a game and therapeutic tool since 2019  \\
\textbf{P13} & 36 / Male & White & Autism; PTSD & AltRoom; Jaunt; Janus; VRChat; Rec Room (avoids Horizon Worlds) & “Used to be a lot more often, not often anymore” due to “demographic changes” \\
\textbf{P14} & 23 / Male & African American & Arthritis, hearing disability & VRChat; Rec Room & Couple of times a week \\
\textbf{P15} & 32 / Female & African American & ADHD, anxiety, depression & VRChat; ChilloutVR & Used to be 2-3 hours every day, slightly less often now \\
\bottomrule
\end{tabular}
\end{center}
\end{table*}

\subsection{Procedure}
We conducted semi-structured interviews via video conferencing software (Zoom). As part of the interview, we engaged participants in a creative activity on a collaborative whiteboard application (Google Jamboard), which allowed participants to draw and paint in multiple colors, import or link images from a device or online searches, type ideas on sticky notes or in text boxes, and create shapes. All interviews were conducted in English and lasted approximately 60 minutes. We provide our full interview protocol in Supplementary Materials.

We began the interview by asking demographic questions. We then asked about their uses for social VR, existing avatar representations in social VR, and their current use of avatars as vessels for disclosure (or not). For example, we asked: “Do you represent or conceal your invisible disability in social VR in any way? How similar is your avatar representation to your physical self?” After establishing the similarities and differences between participants’ virtual and non-virtual selves, we asked if participants were interested in representing their invisible disability / disabilities in social VR. Regardless of their responses, we then discussed how context might impact their decisions to represent (or not represent) their invisible disabilities. We wrapped up the first section by exploring participants’ avatar customization preferences.

We then encouraged participants to generate creative avatar ideas through brainstorming their “ideal avatar representation” in social VR. Participants shared their thought processes while adding images, drawings, or text to a virtual whiteboard to create visual representations of their desired avatar features. Upon completion, we asked participants to describe important features that were not currently available and outline their ideal customization process. As verbal brainstorming often privileges people who are comfortable with and able to generate ideas and express themselves through voice, we designed this activity to offer an alternative opportunity for self-expression and accommodate a broader range of communication styles.

We then explored the importance of context in virtual and non-virtual disclosure. We asked questions such as: “In the non-virtual world, how do you navigate disability disclosure? Would you consider your avatar representation to be the same or different in different settings? How does this compare to other platforms with avatar options, such as Snapchat?” Finally, we discussed the practical implications of incorporating invisible disability representation features into avatars. 

\subsection{Analysis}
We audio recorded and transcribed all interviews, then used open coding to analyze the interviews, drawing on a grounded theory approach \cite{adams2008qualitative}. Three researchers divided the first three transcripts so that each was double-coded. We discussed code discrepancies, created the preliminary codebook, and then split up the remaining transcripts for individual coding. Throughout this process, we iterated upon and refined the codebook. 

After coding, we conducted thematic analysis \cite{braun2012thematic} to identify overarching takeaways and patterns across participants. Our themes were grounded in the interview codes as well as participants’ suggestions during the whiteboard component, which included image, drawing, and textual artifacts. We developed the codebook by reflecting on the proposed themes’ connections to our research questions, comparing areas of overlap and divergence, and reflexively confronting potential analytical assumptions. After multiple rounds of iteration, the co-first authors determined the final thematic findings in consultation with the rest of the research team.

\subsection{Positionality}
Multiple members of our research team have invisible disabilities, and one author uses a feminist disability studies lens in her work. The research team also includes social VR users, accessibility researchers, and people who have grappled with disability disclosure in their own lives. We believe avatar customization for disability representation is an important step toward improving the inclusiveness of social VR spaces, and we aim to include the voices of invisibly disabled VR users in ongoing conversations about fostering positive representation and cultivating diverse spaces. 

\section{Findings}
In this section, we present participants’ uses of social VR, share perceptions of the unique affordances of embodied avatars, detail current and desired invisible disability representations for social VR avatars, and explore disclosure preferences across different social VR contexts.

\subsection{VR as a Social Environment}
Social VR shares many of the same norms and uses as non-virtual social spaces. Participants described using social VR for a variety of purposes, including entertainment (e.g., playing games), socialization (e.g., chatting with others), and professional engagements (e.g., for work). Multiple participants attended support groups, celebrating birthdays, and watching movies in social VR. 

Additionally, for some, virtual platforms provided them essential access to environments when their invisible disabilities or other factors prevented them from leaving home. P9 initially started as a VR developer, but \textit{“when COVID happened, my situation switched from engaging in VR as a platform to build and code on, to a platform to socialize and meet people.”} P3 also described how social VR \textit{“allows me to do things that I can’t do in my physical world, that I used to do in my physical world”}:

\begin{quote}
“In VR, what the person next to me can do, I can also do. But my reality is not that… What I’m getting out of VR is a little different than what most people get out of VR. They get in there, and they want to [play games]. I’m there so that the world doesn’t close in on me… to keep a tiny little porthole of the world open to me. Even though I cannot physically be in it, I can virtually be in it. I can visit a museum through VR. I went up [to] the space station through VR. I’ve gone scuba diving in VR. And the social aspects \textemdash{} I’ve met friends who have been my friends for years, but we live in different states. And we’ve come together.” (P3, age: 56, gender: female) 
\end{quote}

\subsection{Embodied Avatars are Dynamic, Contextual, and Multimodal}
Participants identified key differences in their perceptions of non-embodied and embodied avatars. In general, they viewed non-embodied avatars, such as Facebook profile pictures, Facebook avatars, and Snapchat Bitmojis, as a snapshot of their identity. On the other hand, they perceived embodied avatars as dynamic and authentic representations of themselves. Participants’ current and desired preferences for disability representations reflected their appreciation for multimodal self-expression in social VR. 

\subsubsection{Differences Between Embodied and Non-Embodied Avatars}
Though participants used social media avatars to represent themselves, they felt that the static nature made it difficult for them to convey their current identity. For example, P1 felt that her Facebook avatar was not inclusive of her body size, saying that \textit{“doesn’t look like me at all. I hate it.”} P3 also shared how she used Facebook and social VR for different purposes, and for different audiences, which influenced her degree of comfort with representing herself in embodied and non-embodied contexts.

\begin{quote}
“With Facebook, of course my avatar is a photo of me. But it’s a photo of me 20 years ago. So my avatar doesn’t age there… and I have not changed it to be what I am now. Because the people that are on my Facebook, many of them are people that know me in this physical reality. … I cannot pretend in Facebook that I’m well or not well… I don’t represent a cause, I don’t represent a disease, I don’t represent anything but myself. … Whereas in VR, I do change those things. … [in] virtual reality, there are people there that know and love me, but most people are acquaintances or strangers. So my interaction with them is going to be different [from] my interaction on Facebook.” (P3, 56F)
\end{quote}

Others also felt that embodiment encouraged them to consider the avatar’s full body appearance. For example, P14 mentioned that the \textit{“immersive experience [of being] around a lot of people [and] seeing the avatar move around”} made him want to represent his physical appearance in greater detail on a social VR avatar as opposed to a static avatar on Facebook. P4 used an embodied cat avatar but shared an anecdote about the inconvenience of having her avatar’s eye line towards the ground when engaging with human-sized avatars. 

The multiple, richer communication modes in social VR were also valued by participants. P13 found the inclusion of voice and visuals in social VR to be easier to process: \textit{“[digital is] basically reading text messages, [whereas in social VR you have] voice chat, digital avatar, [or] motion capture.”} P15 mentioned how some of her friends with social anxiety used VRChat sign language or \textit{“an avatar that, when they text and write stuff out, you hear [it] in a different voice that’s not their own”} to engage in a way that was accessible to them. 

Some participants highlighted the heightened interactivity and expressiveness of embodied avatars. For example, as a content creator, P10 wanted to use her social VR avatar in her videos to merge her representations in virtual and non-virtual domains, stating that she \textit{“would actually love using an avatar instead of my ‘real’ body in my YouTube videos”} and represent herself with features such as tails and wings. When discussing the distinctions between social media and social VR, P9 summarized how his interactions differed between the two platforms and emphasized that the immersive affordances of social VR avatars enabled him to better express himself and understand others.

\begin{quote}
“[VR] is an order of magnitude above anything that you would see on Instagram \textemdash{} [or] any other website \textemdash{} because those are just two-dimensional static images or videos. [You] can’t interact with photos, but when you’re there in virtual reality, you can see all the details on the avatar showing that this person is expressing themselves. [You can] see their facial expressions, their movements, and their body language. And it’s real time. No other kind of social platform comes even close to VR.” (P9, 28M) 
\end{quote}

\begin{figure*}[htbp]
  \centering
  \includegraphics[width=\linewidth]{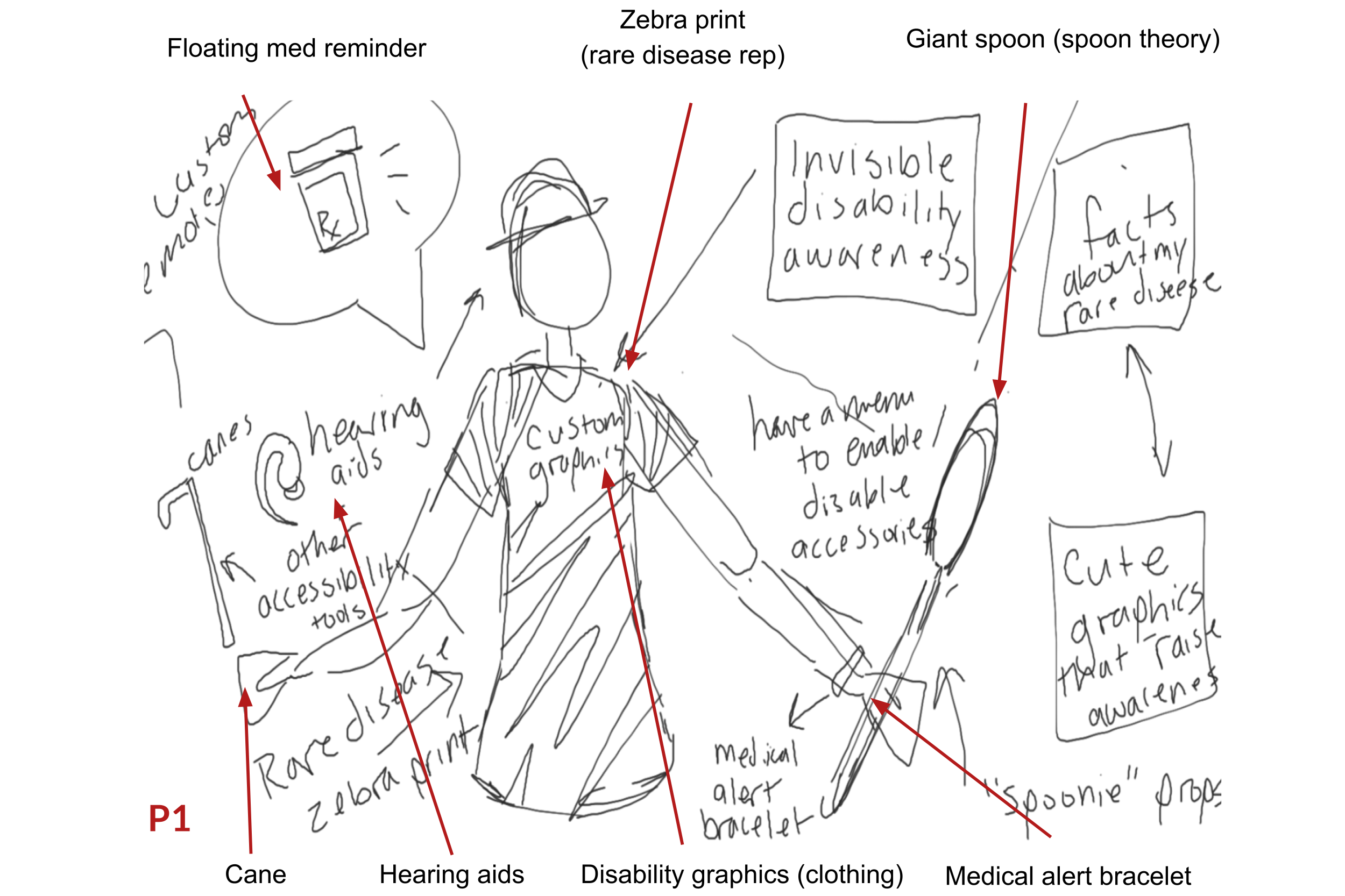}
  \caption{Participant 1’s drawings and suggestions for invisible disability representation during the whiteboard component of the study, featuring a giant spoon to represent spoon theory and other assistive technologies to disclose their disability.} 
  \Description{Participant 1’s whiteboard drawings. At the far left, there is a floating medication reminder inside a speech bubble that represents a floating emote. Below are examples of assistive technology such as a cane and hearing aids. Next is a person’s outline wearing a zebra print beret and shirt to represent Rare Disease Month. The shirt also has a space for a custom disability graphic or a fact about the participant’s rare disease, both aimed at spreading invisible disability awareness. The person also wears a medical alert bracelet and carries a human-sized spoon to represent desired “spoonie” props that point to identification with spoon theory. Finally, there is a note saying “have a menu to enable / disable accessories.”}
  \label{fig:P1}
\end{figure*}

\subsubsection{Multimodal Affordances Supported Disability Representation}
Participants also utilized the multimodal expressiveness of embodied avatars, which manifested primarily through avatar visuals, voices, movements, body language, and facial expressions. For example, participants valued the ability to show dynamic and nuanced facial expressions that reflected their non-virtual self. P9 particularly appreciated that there were many options already available:

\begin{quote}
“I am quite expressive myself, but my invisible disability is nonverbal autism, so I wouldn’t actually say what I’m feeling. I would express it through my avatar’s facial expressions. Sometimes, the number of expressions on an avatar goes as high as 30 or 40 different options. So having many different options to express myself nonverbally is a way that I can get more synchronized and comfortable using an avatar.” (P9, 28M) 
\end{quote}

Similarly, P2 was interested in the possibility of using body language to represent changes in her mood due to bipolar disorder. 

\begin{quote} 
“Let’s say I’m feeling excited or hyper. Your avatar can have a really upbeat energy to it if you’re going to apply an energy to an avatar. Or if you can put in there that you’re feeling slightly depressed, or that you have a low mood or that you’re tired, your avatar would be a little bit lower on the energy… That way my friends could see me and then go like, ‘Oh, that’s kind of the mood she’s in. Let’s cheer her up.’ Or, ‘Let’s just give her her space.’ … They can kind of match their energy to your mood without you having to be so vocal about it.” (P2, 32F) 
\end{quote}

Some participants explained how the embodied movement of VR avatars could represent or support their disability in VR. For example, P12 wanted to utilize either voice or movement: \textit{“if I am going to disclose my disability on the platform, it would be verbally or through using the VR controllers when I’m controlling the hands of the avatar.”} Conversely, though P14’s experiences with arthritis sometimes impacted his mobility in the non-virtual world, he did not want to disclose his disability through movement \textemdash{} he explained how his current avatar reflected his preference for his avatar to move in the same ways as others. Regarding social support, P3 described how embodied avatar interactions could facilitate community empowerment and connection in a disability support group: \textit{“sometimes, our avatars will run over and give somebody a group avatar hug when they’re having a particularly bad time.”} 

\subsection{Current and Desired Invisible Disability Representations}
Throughout the interviews, participants shared how they currently represented or disclosed their invisible disability and brainstormed ways to make avatar customization systems more inclusive. For example, P11 described his avatar to be \textit{“an exaggeration of my physical identity \textemdash{} anonymous, but also close enough to be recognized if you know me personally.”} P12 shared this perspective: \textit{“I try to represent myself as I am, of course, and for me, that involves maybe making an avatar that looks like me but is a little bit cooler if possible.”}

We found that desired methods of representing invisible disabilities in social VR did not always align with non-virtual disclosure practices. Participants desired options for both public (i.e., explicit) and private (i.e., implicit) identity expression. Suggestions for both public and private representations encompassed clothing, accessories, icons, energy bars, and modifications to avatar bodies.

\subsubsection{Public Representations and Expressions of Disability}
Some participants were interested in representing their disabilities in \textbf{public} and explicit ways, meaning that other social VR users could identify their specific disability or disability status from their avatar. 

Eleven participants suggested incorporating customizable clothing, patterns, or accessories that featured disability-related logos. These suggestions often supported participants’ goals of expressing their personal style and educating others about their disability. For example, P1 wanted to have an avatar wearing a zebra print shirt and beret to advocate for Rare Disease Month (Figure \ref{fig:P1}). Others suggested accessories specific to their disability (e.g., P6 wanted a brooch saying “POTS,” P13 proposed a PTSD ribbon, P15 mentioned previously having an ADHD sticker). P4 wished to have an explicit representation for communicating the experience of dyslexia through apparel: 

\begin{quote}
“[On clothing,] have the word dyslexia, but then have it jumbled up so people can see what it looks like when we try to read. Then, they’ll be like, ‘Whoa, that’s not fun.’ And we’re like, ‘Well, welcome to my hell.’ Make it a joke, but also kind of be like, ‘Hey, this is really what it feels like.’” (P4, 38F)
\end{quote}

Other participants wanted to disclose their disabilities by including their assistive technologies (e.g., wheelchairs, canes, crutches, hearing aids), a medical alert bracelet, and \textit{“spoonie”} gear. One example of this was a giant spoon representing Spoon Theory \cite{spoontheory}, a metaphor for the variance in physical or mental energy a disabled person has for daily activities.

Graphics around an avatar could also help with representing someone’s disability in a more subtle, but still public, way. Five participants were interested in representations that floated above their avatars, such as an awareness flag or a dynamic energy bar that correlated with their energy level or “social battery.” For example, P2 proposed \textit{“an energy bar, like in a video game where you have your life meter. If it’s full, you’re full of energy, and it’s all the way green… And if I’m really low on energy, I’ve got a red bar.”} 

\begin{figure}[htbp]
  \centering
  \includegraphics[width=\linewidth]{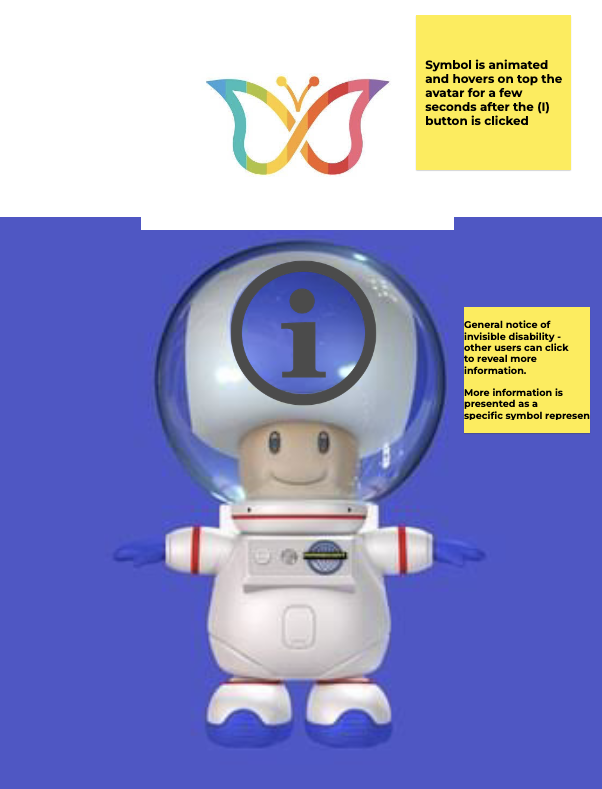}
  \caption{Participant 11’s images, sticky notes, and suggestions for invisible disability representation during the whiteboard component of the study, featuring an information icon and the ADHD rainbow butterfly symbol above the avatar.} 
  \Description{Participant 11’s whiteboard images. There is an image of Toad, an anthropomorphized mushroom character from the Mario franchise, wearing an astronaut suit and helmet. On his cap, there is an information icon (a lowercase i in a circle) and above Toad is the ADHD rainbow butterfly symbol. To Toad’s right, there is a sticky note that reads, “General notice of invisible disability - other users can click to reveal more information. More information is presented as a specific symbol represen-”. To the right of the ADHD symbol, there is a sticky note reading “Symbol is animated and hovers on top the avatar for a few seconds after the (i) button is clicked.”}
  \label{fig:P11}
\end{figure}

While participants specifically criticized iconography they deemed harmful (e.g., puzzle pieces), they valued recognizable icons, such as the rainbow infinity symbol that often represents neurodiversity. When referring to her current avatar’s color-changing hair, P15 shared that it was unintentionally a reference to \textit{“the symbolism, when it comes to the spectrum of neurodiversity \textemdash{} the rainbow… [and] the infinity symbol.”} P11, a designer by trade, did not currently disclose his identity but wished to do so through design patterns and iconography inspired by video games and internet culture (Figure \ref{fig:P11}). He emphasized the transient nature of this representation so as not to greatly disrupt social interactions.

\begin{quote} 
“So ‘i’ is ‘information’ … You click that… [and it pops up] this universal image for neurodiversity. … The image pops up for only a few moments… That way, you can continue your interactions in virtual space without necessarily having the label on you constantly.” (P11, 28M)
\end{quote}

Participants also wanted to use temporary adjustments to indicate their interest in engaging with others. Similar to the energy bar example, P15 described one of her current avatar accessories: \textit{“it would hang on my ear… a [red] microphone with a slash over it. That just tells people, ‘I’m here to vibe, but I can’t talk right now for various reasons.’”} When using this accessory, she also commonly changed her typically colorful avatar into one wearing all black, so as not to \textit{“drive that much attention”} to herself. P11 also proposed a feature that would allow an avatar to turn invisible when a user wanted to avoid social interactions, which he dubbed as a \textit{“stealth mode.”} However, P11 assessed the potential privacy implications of such a feature and concluded that it would be more beneficial to instead turn an avatar translucent; this could convey a similar message of wishing to keep to himself while still alerting other users that there was another user within the space.

\begin{quote} 
“If I’m in [a] VR space and I just want to enjoy myself, and I don’t want anyone coming up to me, I should be able to say that. … [but] if you think about the other users of the space, do you really want an invisible person watching you? … Maybe rather than being invisible… I can just make him more white [translucent] instead… ‘Just don’t come to me. [My avatar’s] not really colorful right now.’” (P11, 28M) 
\end{quote} 

\subsubsection{Private Representations and Expressions of Disability}
Below, we describe participants’ \textbf{private} avatar customizations, which allowed users to represent their invisible disabilities without necessarily disclosing their disabilities to others.

Participants primarily suggested representing their disabilities or experiences through accessories that matched ones they would wear in non-virtual settings. For example, P3 preferred subtle accessories that would more closely resemble non-virtual affordances: \textit{“I would do it with accessories, [but] I don’t want a big sign saying, ‘disabled’ or a different color flashing over here that says ‘invisible illness is no longer invisible.’”}

Others represented themselves through accessories that they did not usually wear in the non-virtual world. During the creative exercise, P8 gave his avatar earbuds, explaining: \textit{“I’m not a big earbud guy. I’m a headphone wearer [in the non-virtual world]… [but] I feel like maybe it would be an extension of trying to conceal what invisible disabilities I would have, just pretending to be normal with stuff like that.”} While this representation of his invisible disability was not explicit to others, he felt that the earbuds were \textit{“more of an expression of [him]”} and preferred to keep this aspect abstract. P14 also utilized headphones and a mask as symbolism for \textit{“blocking out all the noise”} in the environment. Similarly, P9 described how his ideal avatar representation included symbolic protective gear: 

\begin{quote}
“I really want to add knee and elbow pads. If you have ever gone roller skating, you need to put on knee and elbow pads, just in case you fall… The bipolar makes me swing really hard to the left and right. I get these mania phases and these depression phases that have a whiplash to them. And sometimes… I can get hurt or accidentally hurt others. So these pads on my knees and elbows would be a way to symbolize that if I do fall, I gotta make sure I’m protected. I don’t fall too hard, and I can stand up again.” (P9, 28M)
\end{quote}

Four participants currently incorporate implicit alterations into their avatar appearances that privately signal disability. For example, P9 used vibrant colors: \textit{“It’s a way that I can visually differentiate myself from a room full of other people. It’s my way of not just indicating to others that this person’s a little different, but it’s also indicating to myself that I’m a little different.”} P10 also expressed that her avatar’s bushy tail represented her comfort zone and personal bubble. She used the tail as metaphor for her anxiety, further emphasizing the unique affordances of social VR in expanding the bounds of self-expression by allowing fantastical and metaphorical representations of the self. 

\begin{quote} 
“[My avatar’s] tail wraps around it a lot… I like that one because, in some weird way, it gives me a comfort zone of hiding me in a bubble. … Even in real life, people have said I have a bubble, like a huge bubble around me. It’s like you get too close to me… [and] I just back away a little.” (P10, 52F)
\end{quote}

Colors and moving avatar features were helpful for not only representing an invisible disability, but also for managing overwhelming situations through self-stimulating behaviors (i.e., stimming). During the interview, P15 shared her screen when embodying her avatar in VRChat and explained how she designed her avatar to include customizable features and other interactive accessories. For example, she used interactive elements such as a SleightlyBall \cite{sleighball} and a “boop counter.” The “boop counter” was a floating popup tallying the number of people who pat her avatar on the nose, and the SleightlyBall was a customizable ball that she could pass between her avatar’s hands or toss around the VR world. These helped keep herself engaged: \textit{“Sometimes I might play with it [the SleightlyBall] while I’m talking to someone… like a fidget feature. … I can make the sleigh ball slowly move around me, and I can stim.”} She also mentioned that VR mirrors could also help with stimming while still concealing her disability. 

\begin{quote}
“I can put my mirror [on the side], and when I’m a little bit visually overwhelmed… I am constantly watching my hair change colors and my ears change their position, which is why I love avatars that can move. … I’m using my private mirror, so no one knows I’m stimming.” (P15, 32F)
\end{quote}

Though some participants’ brainstormed disability representations were intended to be explicit, they realized that their ideas could still be implicit to other social VR users. For example, while P5 suggested using glowing veins to represent chronic pain from repetitive strain injury, he acknowledged that other users might not understand the symbolism: \textit{“[the glowing veins] still wouldn’t even show pain, because people are gonna look at that and just go like, ‘Cool.’ They just see glowing things on your arm.”}

Others felt that the implicit nature of their representations could still convey some meaning about their disability without disclosing a specific condition. P9 emphasized how his embodied avatar already reflected his mood, especially related to his experiences with Bipolar Type 1. He used VR avatar appearances to communicate his emotions nonverbally, strategically selecting avatars based on his goals for social interactions that day: 

\begin{quote}
“The middle avatar is the avatar I use when I’m in a playful mood. It stands out, but it’s also really friendly looking. I got a lot of people that like to hang around with me in this one if I’m feeling energetic, a little bit extroverted. If we talk about bipolar phases a little bit more, like the manic phase, I’ll use the middle one. … If I’m in a really happy mood, but I pick an avatar with no color, for few emotions, it wouldn’t really match. Same thing vice versa. It also adds some novelty to VRChat… I like to mix it around, so I know there’s always flexibility to iterate and change.” (P9, 28M) 
\end{quote}

Even though most other social VR users would not understand that a disability was being expressed, these private and often implicit disability representations were still valuable to participants.

\subsection{Invisible Disability Disclosure Practices Differ Based on Context}
In this section, we explore participants’ varied motivations for disclosing or sharing their disability identity, informed by social VR environments and the dynamic nature of their disabilities. 

Social factors influenced participants’ disclosure strategies. We found that participants generally belonged to one of three groups: \textbf{Activists}, \textbf{Non-Disclosers}, and \textbf{Situational Disclosers}. \textbf{Activists} were always open to disclosing to spread education and awareness, often tying into advocacy work that the participants did outside of social VR contexts. \textbf{Non-Disclosers} did not want to disclose their invisible disability to others in any circumstances. Lastly, \textbf{Situational Disclosers} were sometimes open to disclosing their disability identities and would rely on context cues to make their disclosure decisions. 

\subsubsection{Activists and Non-Disclosers}
One participant, P1, identified as an \textbf{Activist} and felt comfortable openly sharing her disability in social VR. As an outspoken disability activist, she felt that disclosing her chronic condition was helpful for not only raising awareness but also indicating access needs.
 
\begin{quote}
“I am an activist and an advocate for folks that have invisible disabilities… I’m an open book. So anything that can start a conversation or encourage people to ask questions is really valuable to me, and I feel like having that option is just really important in terms of raising awareness… If I’m going to start a friendship with somebody, I want them to be aware of the fact that I have a chronic illness, because that is a pretty large part of my life, and it does affect my ability to socialize sometimes. Being open about that is important to me personally.” (P1, 32F)
\end{quote}

On the other hand, one participant was not comfortable disclosing in social VR at all. P8, a \textbf{Non-Discloser}, concealed his invisible disability in the non-virtual world and did not want to disclose in social VR either. He felt that others \textit{“wouldn't be very receptive or kind to that”} and he emphasized that his invisible disability was not a core part of his identity, especially as it was \textit{“a condition that didn’t come until the last five years or so, so it’s not like I grew up with it”} (P8). Additionally, he perceived significant stigma against his invisible disability in all contexts, VR or otherwise: \textit{“friends that have also had [my condition], it's like you're treated differently. … I’m trying to hide it so much that, outside of my family, I don’t think I ever opened up to anyone about it”} (P8).

\subsubsection{Situational Disclosers: Disclosure Practices were Informed by Perceptions of Stigma}
Context strongly impacted when and why participants chose to disclose their invisible disabilities. While only one participant identified as an Activist and one as a Non-Discloser, the remaining 13 participants were Situational Disclosers who expressed varying degrees of excitement and hesitancy to represent their disability. Most did not want to disclose in large group contexts because of their perceptions of the stigma they might face and due to prior negative experiences with disclosure. For example, P9 mentioned how he did not actively disclose when interacting with others in social VR: \textit{“I don’t really represent my disability too much, but I also don’t conceal it.”}

For seven Situational Disclosers, their intended audience significantly influenced their avatar design choices, and they often chose to disclose to support other disabled people in a safe space. P15 shared that she often liked to \textit{“people-watch when [entering] a new space”} to identify if it would be a safe space to disclose her disability or race. Most stated that they were more likely to disclose in a disability-specific community rather than in a large social situation. P9 specifically mentioned that disclosing in this context allowed him to empathize and engage; similarly, P2 felt that describing her experience with bipolar disorder in her profile could provide support to others. P4 acknowledged that she did not want to disclose her disability to just anyone in social VR: \textit{“we would mostly have to become friends outside of the virtual reality world for me to feel comfortable to disclose it.”}

Regardless of the scale of a social context, some felt that certain disabilities carried more stigma than others. For example, P4 was comfortable representing dyslexia and chronic pain, but not epilepsy: \textit{“I’ve had more hatred for [having epilepsy] than dyslexia because most people understand dyslexia.”} Furthermore, P5 mentioned that Asperger’s and ADHD were \textit{“poked fun at a lot online,”} and was concerned about being disrespected due to the \textit{“negative public perception”} of his disabilities. P12 also described his prior experiences of facing discrimination against neurodivergent people: 

\begin{quote}
“I don't go out of my way all the time in VR to disclose my disability. Some, if not many, people may not fully understand what the disability is about. Or, they'll jump to conclusions and assume… I don't want to get labeled that way. I don't want to be labeled as someone who's [r-word, slur for people with intellectual disabilities]. … That word has been thrown around a lot recently.” (P12, 31M)
\end{quote}

Given the fluctuating nature of some invisible disabilities, some participants’ disclosure preferences were also dynamic. For example, P2, P6, and P10 chose to disclose based on their mood or flare-ups. P2 would sometimes disclose her disability to help others \textit{“understand what I’m going through [and] to avoid conflict,”} especially in a work setting. P6 also noted that \textit{“I really only disclose if I feel like I’m about to have a flare up.”} Similarly, P10 shared: \textit{“when my Hashimoto’s flares up, I lose lots of hair, I have no energy, and I am just tired… People [can understand], ‘Okay, she’s dealing with that today.’”} P3 elaborated on how having easily customizable avatars played a psychological role in grounding her VR experiences: \textit{“[having options] gives you virtual reality with just a little bit more normalcy. … It gives you a little bit of an anchor to not overwhelm you”} (P3).

Participants also shared how they wished to toggle specific avatar features on and off to support their dynamic disclosure preferences. P9 felt comfortable and open to disclosing his disability in social VR, but mentioned that having a toggle for disclosure features could be crucial for avoiding ableism in unwelcoming settings. As someone with social anxiety, P10 also thought that having options to both (1) have some disability representations permanently fixed to her avatar and (2) select other disability representations on a daily basis would be most helpful: \textit{“if you have social anxiety every single day, there’s no reason to switch that on and off. That gives you anxiety, having to go do that every day.”} P3 explained how the toggles could be used for both representing and managing her invisible disability: 

\begin{quote}
“My VR avatar is whatever I’m feeling at the moment. If I’m feeling immersed in my invisible illness, then I’m probably gonna want my avatar to be also. If I’m feeling like, ‘okay, I’m escaping reality,’ I don’t want my avatar to be even close to what I am. [In our social VR disability support group,] there will still be some days that I don’t want to manifest [my disability]. And then other days where it’s like, ‘Yeah, I’m feeling like crap and I’ll let you people know it in a nonverbal way,’ and I will come in there all crippled up with my avatar, altered to represent how I feel right now simply because I am trying to communicate it in a nonverbal way.” (P3, 56F)
\end{quote}

\subsubsection{Representing and Disclosing Intersectional Identities}
Disability is one facet of the complex human experience; other intersecting identities like gender and race may also impact disclosure. Some participants described their experiences as people with multiple marginalized identities. For example, P15 started creating custom avatars because existing options were not inclusive: \textit{“I got so sick of really cool avatars just being white.”} During the study, she referenced art depicting Black women in diverse aesthetic styles as inspiration for her avatar designs. P6 also acknowledged her intersectional identities of race, gender, and invisible disability, and drew on her daily experiences in social and workplace settings to inform her avatar disclosure preferences:

\begin{quote}
“I feel like I’m seen as unreliable. Obviously, I’m also a Black woman, so that compounds things \textemdash{} that intersection of discrimination. But I do feel like I actively do not get treated well when I talk about it. In social situations, I'm not concerned about it the way that I am in professional circumstances. … The intersection of disability and race has made me a lot less likely to disclose over the years. As I’ve seen, figures of authority are not necessarily empathetic or respectful about it.” (P6, 24F)
\end{quote}

\subsubsection{Perceptions on Safety and Harassment in Social VR}
Some participants perceived social VR to be a relatively safe space where they could freely express their disability-related identities without worrying as much about the stigma they faced in the non-virtual world. For example, P9 felt more open to disclosing disabilities in social VR because the “disability” label was not permanent: \textit{“I feel like once I label myself [with a disability] in real life, it becomes something that kind of follows you around.”}

When speaking about the personal importance of representing his invisible disability, P12 drew parallels between LGBTQ and disability representation in the media. He cited LGBTQ award-winners at the Tony Awards as positive examples of ways that queer people could confidently express and disclose their identity through clothing and other visual means. 

However, others were wary of potentially harmful social VR environments. For example, P13 mentioned a news article reporting about a woman being assaulted in social VR and briefly recalled his own experiences of being assaulted in VR. P1 also shared how disability disclosure in social VR could result in harassment:

\begin{quote}
“If you’re going into more of a rough and rowdy crowd, like in VRChat, you don’t want to make yourself a target for people that might look for things to hurt you with. … If you’re going to a place that you don’t really know very well, I think it would be a good idea to be able to turn [disability representations] off or on.” (P1, 32F) 
\end{quote}

\section{Discussion}
In our study, we found that embodied avatars allowed for greater creative expression than social media avatars. To answer our research questions, we identified different ways and contexts in which invisibly disabled people wished to represent themselves. Unlike people with visible disabilities, who have suggested depicting disabilities through assistive technology on their avatars \cite{zhang2023diary, zhang2022s, mack2023towards}, invisibly disabled people preferred to adopt more nuanced and symbolic approaches such as clothing designs (e.g., custom disability-related graphics) and accessories (e.g., a giant spoon, a floating energy bar). Additionally, depending on the social VR context, participants chose between disclosing through public representations with explicit meaning and private representations with personal significance and implicit meaning. 

\subsection{Differences Between Embodied and Non-Embodied Social Contexts}
We uncovered distinct contrasts between disability representation preferences for embodied avatars compared to digital social media platforms such as Facebook or Snapchat. Social media limits disclosure to images, text, and video, whereas social VR offers users multiple channels for sensory expression (e.g., facial expressions, voice), movement patterns, and wearable disability representation through embodied, full-body avatars. While Mack et al. \cite{mack2023towards} explored contextual differences across digital gaming and social platforms such as Roblox and Snapchat, we extend prior findings on invisible disability representation by examining social VR’s embodied affordances for identity expression. 

One such embodied affordance is avatar movement, also known as locomotion. For people with chronic conditions (e.g., chronic pain), mobility or motor disabilities, or disabilities which involve stimming, VR avatars currently conceal their non-virtual behaviors and locomotion techniques \cite{bozgeyikli2016locomotion, jaeger2013cyber}. While some preferred to have their avatars move similarly to others’ avatars, others wished to represent their non-virtual locomotion techniques and body language. Prior work has examined how diverse representations of locomotion can make video games more inclusive \cite{mason2022including}. Most research on VR and disabilities takes a medicalized approach, focusing primarily on diagnosis, training, or rehabilitation (e.g., \cite{matamala2019immersive, ahmadpour2019virtual, harrison2002role, laufer2011virtual}), with some exceptions (e.g., \cite{zhang2022s, zhang2023diary, mack2023towards}). We encourage researchers to further examine the potential for locomotion to serve as another method of disability disclosure and identity affirmation for people with both visible and invisible disabilities.

Social VR establishes a context distinct from both the non-virtual world and digital social media platforms \cite{freeman2021body}. The relative newness of this social setting brings significant potential, with some referring to VR as “a technological ‘wild west’” \cite{harley2022would}. While social VR allows users to have greater degrees of self-expression and identity affirmation, it also introduces new risks of hate and harassment \cite{freeman2022disturbing, zhang2023diary}. As a result, more research is needed to understand how users navigate the newfound advantages of identity representation within VR while also protecting themselves and mitigating harm.

\subsection{Disability Disclosure Strategies}
Identity representation preferences, including disability disclosure, often vary by context and condition. These preferences become more nuanced given the embodiment and real-time immersion in social VR \cite{freeman2021body, won2023your}. Similar to the three strategic approaches proposed by Furr et al. \cite{furr2016strategic}, we identified three disclosure types that emerged from our findings: \textbf{Activists}, \textbf{Situational Disclosers}, and \textbf{Non-Disclosers}. The \textbf{Activist} participant was always open to disclosing to spread education and awareness. The \textbf{Situational Disclosers} were sometimes open to disclosing and relied on context cues to make their disclosure decisions. The \textbf{Non-Discloser} participant preferred to conceal his disability at all times. 

Beyond \textit{when} or \textit{why} people would disclose their invisible disability, we also uncovered insights into \textit{how} invisibly disabled people preferred to represent themselves. Social VR users sometimes used \textbf{public} representations of their disability to explicitly share their identity, express their personal style, and raise awareness about their disability. Others preferred using \textbf{private} representations to still represent their invisible disability and affirm their identity without necessarily disclosing to other social VR users. This binary framework points specifically to nuances surrounding embodied invisible disability expression; in contrast to people with visible disabilities \cite{zhang2022s, mack2023towards}, people with invisible disabilities have the unique experiences of needing to choose to actively represent their disability in both non-virtual and virtual settings. 

Our work differs from Furr et al.’s framework \cite{furr2016strategic}. For people with visible disabilities adopting the \textbf{open disclosure approach} on social media, Furr et al. explained that this approach \textit{“involves blunt, uncensored, unapologetic, and upfront representations of the disability on an open social media platform… [and] often reveals complete details about the nature of the physical disability and is sometimes accompanied by humor or proclamations of strength.”} In our framework, however, we highlight that activism is inherent to open invisible disability disclosure. Our participants practiced activism in many ways: educating others about a disability, addressing invalidating comments about not being disabled enough, or explaining their fluctuating daily experiences and access needs. This builds on findings from Mack et al. \cite{mack2023towards} about the importance of using avatars for increasing disability awareness and education.

Furthermore, Furr et al.’s \textbf{secure approach} does not map directly onto the social VR experiences we uncovered in our study. There were some similarities \textemdash{} for example, we also found that Situational Disclosers in VR had prior negative disclosure experiences and \textit{“routinely disclose[d] more comprehensive information among a secure population of persons with similar disabilities”} \cite{furr2016strategic}, such as in a support group setting. However, one primary distinction is that users are more likely to encounter and engage with strangers in social VR than they are on social media \cite{freeman2021hugging}. While social media users typically chose to only disclose to audiences deemed safe and supportive, Situational Disclosers in social VR leveraged different methods to disclose in public spaces with strangers. To minimize risks of harm, they sometimes utilized implicit features, such as earbuds or elbow and knee pads on their embodied avatar, to represent and reaffirm their identity and communicate access needs while still obscuring their invisible disability to a broad audience.

In our work, we build on prior research examining disability disclosure on social media platforms. We present three types of disclosure, and two methods to disclose, that capture the nuances associated with representing invisible disabilities through embodied avatars in social VR. 

\subsection{Design Recommendations for Inclusive Embodied Avatars} 
Based on our findings, we offer the following design recommendations for invisible disability representation in avatars. 

\begin{enumerate}
\item \textbf{Create more options for multimodal customization and interaction.} 
    \begin{itemize}
      \item For some invisibly disabled people, certain forms of communication are more accessible than others. We recommend further developing options to communicate verbally (i.e., voice \cite{povinelli2024springboard, kao2022audio, zhang2021social}), through sign (i.e., high fidelity hand tracking \cite{dzardanova2022sign}), visually (i.e., sketching \cite{drey2020vrsketchin}), etc. 
      \item Communication through movement can be valuable for representing one’s disability. Platforms should allow users to customize their movements and facial expressions for easier and more authentic representation. 
    \end{itemize}
\item \textbf{Include presets for both implicit and explicit options for invisible disability representation on social VR customization platforms.} 
    \begin{itemize}
      \item Even though some participants suggested highly individualized representation ideas, \textbf{we recommend including awareness-building apparel and accessory options in customization libraries} (e.g., zebra print for Rare Disease Month, dyslexia-inspired T-shirt design, giant spoon accessory for representing chronic conditions \cite{spoontheory}). 
      \item Navigating customization can be challenging for multiple reasons, so offering a number of preset invisible disability representation features can help with boosting awareness and facilitating self-expression. 
    \end{itemize}
\item \textbf{Support dynamic disclosure preferences and fluctuating disability experiences.}
    \begin{itemize}
      \item Many invisibly disabled people have \textbf{varied experiences from day-to-day}. Users should be able to change their avatar on a regular basis to reflect their energy level (or “number of spoons”) that day. 
      \item To allow users the utmost autonomy in their avatar disclosure decisions, platforms should offer an \textbf{easy-to-use toggle} to turn disability representation features on or off.
       \item Certain customizations, such as energy level, may need to be set every time a user logs on to a platform. We recommend \textbf{creating a pop up that reminds users to set this value} when they first enter a social VR space so they can express their disability-related identity without needing to search for additional steps. Other representations that may be less fluctuating, such as clothing, can be included in a menu that users can access anytime when in a social VR space. To reduce cognitive load, these less frequently accessed options do not need to constantly be visible.
     \item Users may also have varied comfort levels with disclosure in different social VR spaces, but \textbf{constant manual updates to avatars can be costly in terms of time, effort, and / or capital}. To streamline this process, if a user enters a disability-specific social VR space, their clothing-based or fixed disability representations could automatically turn on since the system recognizes that they are in a safe space. On the other hand, when entering a more public space such as a VRChat lobby, the user can automatically switch to a non-disclosing avatar to protect against harassment.
    \end{itemize}
\item \textbf{Make the avatar customization process more accessible and flexible.}
    \begin{itemize}
      \item Many social VR users do not have the programming background necessary to manually fine-tune more individualized customization features like those that might represent invisible disabilities. We encourage social VR platforms to \textbf{integrate non-code options} that can lower the barrier to entry for avatar customization. 
      \item Because of the \textbf{personal significance of many suggested disability representations}, we recommend enabling social VR users to upload disability-related customizations (e.g., custom apparel logos). To prevent abuse of this feature, custom designs can be subject to a brief review period and users can report other users who upload graphic, hateful, or otherwise inappropriate designs.
       \item \textbf{Minimize or remove the physical burden of swiping} through long lists of customization options to enhance accessibility, including customization by way of text, voice, or drawing.
       \item \textbf{Multimodal artificial intelligence} could assist with converting avatar descriptions or sketches into avatars to reduce the cognitive load and overhead of learning Unity or Blender to create custom avatar designs. For non-technical users who wish to fine-tune or have more granular control over their avatar’s appearance, we recommend for systems to segment different components of the avatars (e.g., hair color, clothing style, disability-related accessories) to streamline the prompting process and ensure that avatar features that the user is satisfied with do not get regenerated.
    \end{itemize}
\end{enumerate}

\subsection{Limitations and Future Work}
We acknowledge limitations to our study recruitment. Participants needed to feel comfortable disclosing they had an invisible disability to participate in our study; as such, we did not capture the perspectives of invisibly disabled people who do not wish to disclose their disabilities at all. We also considered that some people with chronic conditions, neurodivergence, or other invisible disabilities might not identify as invisibly disabled, so we included examples of invisible disabilities in our recruitment message. Additionally, as we recruited adults aged 18 or older, we did not connect with the large population of youth VR users and gamers. Nonetheless, our participants shared diverse viewpoints, including perspectives on how their invisible disability intersected with other facets of their identity, while also having overlapping experiences. 

Given that people with invisible disabilities comprise a diverse set of users, there are disabilities that are not represented amongst our participants. We do not aim to generalize across the wants and needs of all invisibly disabled and / or neurodivergent people. Rather, we aim to start a conversation around the importance of working with these communities to understand what they value to achieve inclusion in social VR.

Though we recruited from organizations, email lists, and social media threads globally, many of our participants were located in the United States. We recognize the geographic limitations of this study and acknowledge the importance of situating findings in diverse cultural and socioeconomic contexts. Disability representation symbols are culturally and geographically dependent. For example, while symbols like a spoon \cite{spoontheory} are globally recognized as a representation of and metaphor for invisible disability, others, like the hidden disabilities sunflower \cite{sunflower}, are more regional. We encourage future work to investigate invisible disability representation preferences across cultural and geographic contexts.

Lastly, to some degree, it is difficult to disentangle precisely \textit{what} influences invisibly disabled people’s disclosure preferences. In our study, we did not explicitly determine how the \textbf{audience} (e.g., primarily strangers in VR vs. primarily acquaintances and friends on social media), the \textbf{culture of a platform} (e.g., sharing synchronous experiences in social VR vs. sharing periodic life updates on social media), or the \textbf{sense of embodiment} (much higher in VR than on social media) can influence a person’s disclosure preferences. Future work could examine the varied contexts in social VR to more explicitly identify how avatar representation preferences differ between social VR and social media.

\section{Conclusion}
Avatar customization options determine who gets to express what identity; however, there are currently few opportunities for people to represent their invisible disabilities. In this study, we found that invisibly disabled people wished to convey their disability identities through VR’s multimodal affordances (e.g., facial expressions, body language) and avatar features or accessories (e.g., clothing, icons). Many preferred to toggle disability representations on and off in different contexts. We defined a framework distinguishing between public and private representations and identified three disclosure types: Activists, Non-Disclosers, and Situational Disclosers. To our knowledge, this is the first study to specifically examine the embodied avatar representation preferences of invisibly disabled social VR users. Our findings contribute to the ongoing conversation of designing with and for people with invisible disabilities.

\begin{acks}
We thank our anonymous reviewers for their feedback and our participants for their insights. This work was partially supported by a gift from Meta (Meta Platforms, Inc.).
\end{acks}

%%
%% The next two lines define the bibliography style to be used, and
%% the bibliography file.
\bibliographystyle{ACM-Reference-Format}
\bibliography{references}

%%% -*-BibTeX-*-
%%% Do NOT edit. File created by BibTeX with style
%%% ACM-Reference-Format-Journals [18-Jan-2012].

\begin{thebibliography}{71}

%%% ====================================================================
%%% NOTE TO THE USER: you can override these defaults by providing
%%% customized versions of any of these macros before the \bibliography
%%% command.  Each of them MUST provide its own final punctuation,
%%% except for \shownote{}, \showDOI{}, and \showURL{}.  The latter two
%%% do not use final punctuation, in order to avoid confusing it with
%%% the Web address.
%%%
%%% To suppress output of a particular field, define its macro to expand
%%% to an empty string, or better, \unskip, like this:
%%%
%%% \newcommand{\showDOI}[1]{\unskip}   % LaTeX syntax
%%%
%%% \def \showDOI #1{\unskip}           % plain TeX syntax
%%%
%%% ====================================================================

\ifx \showCODEN    \undefined \def \showCODEN     #1{\unskip}     \fi
\ifx \showDOI      \undefined \def \showDOI       #1{#1}\fi
\ifx \showISBNx    \undefined \def \showISBNx     #1{\unskip}     \fi
\ifx \showISBNxiii \undefined \def \showISBNxiii  #1{\unskip}     \fi
\ifx \showISSN     \undefined \def \showISSN      #1{\unskip}     \fi
\ifx \showLCCN     \undefined \def \showLCCN      #1{\unskip}     \fi
\ifx \shownote     \undefined \def \shownote      #1{#1}          \fi
\ifx \showarticletitle \undefined \def \showarticletitle #1{#1}   \fi
\ifx \showURL      \undefined \def \showURL       {\relax}        \fi
% The following commands are used for tagged output and should be
% invisible to TeX
\providecommand\bibfield[2]{#2}
\providecommand\bibinfo[2]{#2}
\providecommand\natexlab[1]{#1}
\providecommand\showeprint[2][]{arXiv:#2}

\bibitem[Abramczuk et~al\mbox{.}(2023)]%
        {abramczuk2023meet}
\bibfield{author}{\bibinfo{person}{Katarzyna Abramczuk}, \bibinfo{person}{Zbigniew Bohdanowicz}, \bibinfo{person}{Bartosz Muczyński}, \bibinfo{person}{Kinga~H. Skorupska}, {and} \bibinfo{person}{Daniel Cnotkowski}.} \bibinfo{year}{2023}\natexlab{}.
\newblock \showarticletitle{Meet Me in VR! Can VR Space Help Remote Teams Connect: A Seven-Week Study With Horizon Workrooms}.
\newblock \bibinfo{journal}{\emph{International Journal of Human-Computer Studies}}  \bibinfo{volume}{179} (\bibinfo{year}{2023}), \bibinfo{pages}{103--104}.
\newblock
\showISSN{1071-5819}
\urldef\tempurl%
\url{https://doi.org/10.1016/j.ijhcs.2023.103104}
\showDOI{\tempurl}


\bibitem[Adams et~al\mbox{.}(2008)]%
        {adams2008qualitative}
\bibfield{author}{\bibinfo{person}{Anne Adams}, \bibinfo{person}{Peter Lunt}, {and} \bibinfo{person}{Paul Cairns}.} \bibinfo{year}{2008}\natexlab{}.
\newblock \showarticletitle{A qualitative approach to HCI research}.
\newblock  (\bibinfo{year}{2008}).
\newblock
\urldef\tempurl%
\url{https://oro.open.ac.uk/11911/}
\showURL{%
\tempurl}


\bibitem[Ahmadpour et~al\mbox{.}(2019)]%
        {ahmadpour2019virtual}
\bibfield{author}{\bibinfo{person}{Naseem Ahmadpour}, \bibinfo{person}{Hayden Randall}, \bibinfo{person}{Harsham Choksi}, \bibinfo{person}{Antony Gao}, \bibinfo{person}{Christopher Vaughan}, {and} \bibinfo{person}{Philip Poronnik}.} \bibinfo{year}{2019}\natexlab{}.
\newblock \showarticletitle{Virtual Reality interventions for acute and chronic pain management}.
\newblock \bibinfo{journal}{\emph{The international journal of biochemistry \& cell biology}}  \bibinfo{volume}{114} (\bibinfo{year}{2019}), \bibinfo{pages}{105568}.
\newblock
\urldef\tempurl%
\url{https://doi.org/10.1016/j.biocel.2019.105568}
\showURL{%
\tempurl}


\bibitem[Blackwell et~al\mbox{.}(2019)]%
        {blackwell2019harassment}
\bibfield{author}{\bibinfo{person}{Lindsay Blackwell}, \bibinfo{person}{Nicole Ellison}, \bibinfo{person}{Natasha Elliott-Deflo}, {and} \bibinfo{person}{Raz Schwartz}.} \bibinfo{year}{2019}\natexlab{}.
\newblock \showarticletitle{Harassment in Social VR: Implications for Design}. In \bibinfo{booktitle}{\emph{2019 IEEE Conference on Virtual Reality and 3D User Interfaces (VR)}}. \bibinfo{pages}{854--855}.
\newblock
\urldef\tempurl%
\url{https://doi.org/10.1109/VR.2019.8798165}
\showDOI{\tempurl}


\bibitem[Boyd et~al\mbox{.}(2018)]%
        {boyd2018leveling}
\bibfield{author}{\bibinfo{person}{Louanne Boyd}, \bibinfo{person}{Kendra Day}, \bibinfo{person}{Natalia Stewart}, \bibinfo{person}{Kaitlyn Abdo}, \bibinfo{person}{kathleen lamkin}, {and} \bibinfo{person}{Erik Linstead}.} \bibinfo{year}{2018}\natexlab{}.
\newblock \showarticletitle{Leveling the Playing Field: Supporting Neurodiversity Via Virtual Realities}.
\newblock \bibinfo{journal}{\emph{Technology \& Innovation}}  \bibinfo{volume}{20} (\bibinfo{date}{11} \bibinfo{year}{2018}).
\newblock
\urldef\tempurl%
\url{https://doi.org/10.21300/20.1-2.2018.105}
\showDOI{\tempurl}


\bibitem[Bozgeyikli et~al\mbox{.}(2016)]%
        {bozgeyikli2016locomotion}
\bibfield{author}{\bibinfo{person}{Evren Bozgeyikli}, \bibinfo{person}{Andrew Raij}, \bibinfo{person}{Srinivas Katkoori}, {and} \bibinfo{person}{Rajiv Dubey}.} \bibinfo{year}{2016}\natexlab{}.
\newblock \showarticletitle{Locomotion in virtual reality for individuals with autism spectrum disorder}. In \bibinfo{booktitle}{\emph{Proceedings of the 2016 symposium on spatial user interaction}}. \bibinfo{pages}{33--42}.
\newblock
\urldef\tempurl%
\url{https://doi.org/10.1145/2983310.2985763}
\showURL{%
\tempurl}


\bibitem[Braun and Clarke(2006)]%
        {braun2006using}
\bibfield{author}{\bibinfo{person}{Virginia Braun} {and} \bibinfo{person}{Victoria Clarke}.} \bibinfo{year}{2006}\natexlab{}.
\newblock \showarticletitle{Using thematic analysis in psychology}.
\newblock \bibinfo{journal}{\emph{Qualitative research in psychology}} \bibinfo{volume}{3}, \bibinfo{number}{2} (\bibinfo{year}{2006}), \bibinfo{pages}{77--101}.
\newblock
\urldef\tempurl%
\url{https://doi.org/10.1191/1478088706qp063oa}
\showURL{%
\tempurl}


\bibitem[Braun and Clarke(2012)]%
        {braun2012thematic}
\bibfield{author}{\bibinfo{person}{Virginia Braun} {and} \bibinfo{person}{Victoria Clarke}.} \bibinfo{year}{2012}\natexlab{}.
\newblock \bibinfo{booktitle}{\emph{Thematic analysis}}.
\newblock \bibinfo{publisher}{American Psychological Association}.
\newblock
\urldef\tempurl%
\url{https://doi.org/10.1080/17439760.2016.1262613}
\showURL{%
\tempurl}


\bibitem[Caserman et~al\mbox{.}(2019)]%
        {caserman2019real}
\bibfield{author}{\bibinfo{person}{Polona Caserman}, \bibinfo{person}{Augusto Garcia-Agundez}, \bibinfo{person}{Robert Konrad}, \bibinfo{person}{Stefan G{\"o}bel}, {and} \bibinfo{person}{Ralf Steinmetz}.} \bibinfo{year}{2019}\natexlab{}.
\newblock \showarticletitle{Real-time body tracking in virtual reality using a Vive tracker}.
\newblock \bibinfo{journal}{\emph{Virtual Reality}}  \bibinfo{volume}{23} (\bibinfo{year}{2019}), \bibinfo{pages}{155--168}.
\newblock
\urldef\tempurl%
\url{https://doi.org/10.1007/s10055-018-0374-z}
\showURL{%
\tempurl}


\bibitem[{Centers for Disease Control and Prevention}(2023)]%
        {cdcdefinition}
\bibfield{author}{\bibinfo{person}{{Centers for Disease Control and Prevention}}.} \bibinfo{year}{2023}\natexlab{}.
\newblock \showarticletitle{Disability Inclusion as a Cornerstone for Health Equity}.
\newblock  (\bibinfo{year}{2023}).
\newblock
\urldef\tempurl%
\url{https://www.cdc.gov/healthequity/features/disability-inclusion/index.html}
\showURL{%
\tempurl}


\bibitem[Dalgin and Bellini(2008)]%
        {dalgin2008invisible}
\bibfield{author}{\bibinfo{person}{Rebecca~Spirito Dalgin} {and} \bibinfo{person}{James Bellini}.} \bibinfo{year}{2008}\natexlab{}.
\newblock \showarticletitle{Invisible Disability Disclosure in an Employment Interview: Impact on Employers' Hiring Decisions and Views of Employability}.
\newblock \bibinfo{journal}{\emph{Rehabilitation Counseling Bulletin}} \bibinfo{volume}{52}, \bibinfo{number}{1} (\bibinfo{year}{2008}), \bibinfo{pages}{6--15}.
\newblock
\urldef\tempurl%
\url{https://doi.org/10.1177/0034355207311311}
\showDOI{\tempurl}


\bibitem[Davis and Chansiri(2019)]%
        {davis2019digital}
\bibfield{author}{\bibinfo{person}{Donna~Z. Davis} {and} \bibinfo{person}{Karikarn Chansiri}.} \bibinfo{year}{2019}\natexlab{}.
\newblock \showarticletitle{Digital Identities -- Overcoming Visual Bias Through Virtual Embodiment}.
\newblock \bibinfo{journal}{\emph{Information, Communication \& Society}} \bibinfo{volume}{22}, \bibinfo{number}{4} (\bibinfo{year}{2019}), \bibinfo{pages}{491--505}.
\newblock
\urldef\tempurl%
\url{https://doi.org/10.1080/1369118X.2018.1548631}
\showDOI{\tempurl}
\showeprint{https://doi.org/10.1080/1369118X.2018.1548631}


\bibitem[Drey et~al\mbox{.}(2020)]%
        {drey2020vrsketchin}
\bibfield{author}{\bibinfo{person}{Tobias Drey}, \bibinfo{person}{Jan Gugenheimer}, \bibinfo{person}{Julian Karlbauer}, \bibinfo{person}{Maximilian Milo}, {and} \bibinfo{person}{Enrico Rukzio}.} \bibinfo{year}{2020}\natexlab{}.
\newblock \showarticletitle{Vrsketchin: Exploring the design space of pen and tablet interaction for 3d sketching in virtual reality}. In \bibinfo{booktitle}{\emph{Proceedings of the 2020 CHI conference on human factors in computing systems}}. \bibinfo{pages}{1--14}.
\newblock
\urldef\tempurl%
\url{https://doi.org/10.1145/3313831.3376628}
\showURL{%
\tempurl}


\bibitem[Dzardanova et~al\mbox{.}(2022)]%
        {dzardanova2022sign}
\bibfield{author}{\bibinfo{person}{Elena Dzardanova}, \bibinfo{person}{Vlasios Kasapakis}, \bibinfo{person}{Spyros Vosinakis}, {and} \bibinfo{person}{Konstantina Psarrou}.} \bibinfo{year}{2022}\natexlab{}.
\newblock \showarticletitle{Sign Language in Immersive VR: Design, Development, and Evaluation of a Testbed Prototype}. In \bibinfo{booktitle}{\emph{Proceedings of the 28th ACM Symposium on Virtual Reality Software and Technology}}. \bibinfo{pages}{1--2}.
\newblock
\urldef\tempurl%
\url{https://doi.org/10.1145/3562939.3565676}
\showURL{%
\tempurl}


\bibitem[Evans(2019)]%
        {evans2019trial}
\bibfield{author}{\bibinfo{person}{Heather~D. Evans}.} \bibinfo{year}{2019}\natexlab{}.
\newblock \showarticletitle{‘Trial by Fire’: Forms of Impairment Disclosure and Implications for Disability Identity}.
\newblock \bibinfo{journal}{\emph{Disability \& Society}} \bibinfo{volume}{34}, \bibinfo{number}{5} (\bibinfo{year}{2019}), \bibinfo{pages}{726--746}.
\newblock
\urldef\tempurl%
\url{https://doi.org/10.1080/09687599.2019.1580187}
\showDOI{\tempurl}


\bibitem[Faucett et~al\mbox{.}(2017)]%
        {faucett2017visibility}
\bibfield{author}{\bibinfo{person}{Heather~A Faucett}, \bibinfo{person}{Kate~E Ringland}, \bibinfo{person}{Amanda~LL Cullen}, {and} \bibinfo{person}{Gillian~R Hayes}.} \bibinfo{year}{2017}\natexlab{}.
\newblock \showarticletitle{(In) visibility in disability and assistive technology}.
\newblock \bibinfo{journal}{\emph{ACM Transactions on Accessible Computing (TACCESS)}} \bibinfo{volume}{10}, \bibinfo{number}{4} (\bibinfo{year}{2017}), \bibinfo{pages}{1--17}.
\newblock
\urldef\tempurl%
\url{https://doi.org/10.1145/3132040}
\showURL{%
\tempurl}


\bibitem[Freeman and Acena(2021)]%
        {freeman2021hugging}
\bibfield{author}{\bibinfo{person}{Guo Freeman} {and} \bibinfo{person}{Dane Acena}.} \bibinfo{year}{2021}\natexlab{}.
\newblock \showarticletitle{Hugging from a distance: Building interpersonal relationships in social virtual reality}. In \bibinfo{booktitle}{\emph{Proceedings of the 2021 ACM International Conference on Interactive Media Experiences}}. \bibinfo{pages}{84--95}.
\newblock
\urldef\tempurl%
\url{https://doi.org/10.1145/3452918.3458805}
\showURL{%
\tempurl}


\bibitem[Freeman and Acena(2022)]%
        {freeman2022acting}
\bibfield{author}{\bibinfo{person}{Guo Freeman} {and} \bibinfo{person}{Dane Acena}.} \bibinfo{year}{2022}\natexlab{}.
\newblock \showarticletitle{"Acting Out" Queer Identity: The Embodied Visibility in Social Virtual Reality}.
\newblock \bibinfo{journal}{\emph{Proc. ACM Hum.-Comput. Interact.}} \bibinfo{volume}{6}, \bibinfo{number}{CSCW2}, Article \bibinfo{articleno}{263} (\bibinfo{date}{nov} \bibinfo{year}{2022}), \bibinfo{numpages}{32}~pages.
\newblock
\urldef\tempurl%
\url{https://doi.org/10.1145/3555153}
\showDOI{\tempurl}


\bibitem[Freeman and Maloney(2021)]%
        {freeman2021body}
\bibfield{author}{\bibinfo{person}{Guo Freeman} {and} \bibinfo{person}{Divine Maloney}.} \bibinfo{year}{2021}\natexlab{}.
\newblock \showarticletitle{Body, Avatar, and Me: The Presentation and Perception of Self in Social Virtual Reality}.
\newblock \bibinfo{journal}{\emph{Proc. ACM Hum.-Comput. Interact.}} \bibinfo{volume}{4}, \bibinfo{number}{CSCW3}, Article \bibinfo{articleno}{239} (\bibinfo{date}{jan} \bibinfo{year}{2021}), \bibinfo{numpages}{27}~pages.
\newblock
\urldef\tempurl%
\url{https://doi.org/10.1145/3432938}
\showDOI{\tempurl}


\bibitem[Freeman et~al\mbox{.}(2022a)]%
        {freeman2022re}
\bibfield{author}{\bibinfo{person}{Guo Freeman}, \bibinfo{person}{Divine Maloney}, \bibinfo{person}{Dane Acena}, {and} \bibinfo{person}{Catherine Barwulor}.} \bibinfo{year}{2022}\natexlab{a}.
\newblock \showarticletitle{(Re)Discovering the Physical Body Online: Strategies and Challenges to Approach Non-Cisgender Identity in Social Virtual Reality}. In \bibinfo{booktitle}{\emph{Proceedings of the 2022 CHI Conference on Human Factors in Computing Systems}} (New Orleans, LA, USA) \emph{(\bibinfo{series}{CHI '22})}. \bibinfo{publisher}{Association for Computing Machinery}, \bibinfo{address}{New York, NY, USA}, Article \bibinfo{articleno}{118}, \bibinfo{numpages}{15}~pages.
\newblock
\showISBNx{9781450391573}
\urldef\tempurl%
\url{https://doi.org/10.1145/3491102.3502082}
\showDOI{\tempurl}


\bibitem[Freeman et~al\mbox{.}(2022b)]%
        {freeman2022disturbing}
\bibfield{author}{\bibinfo{person}{Guo Freeman}, \bibinfo{person}{Samaneh Zamanifard}, \bibinfo{person}{Divine Maloney}, {and} \bibinfo{person}{Dane Acena}.} \bibinfo{year}{2022}\natexlab{b}.
\newblock \showarticletitle{Disturbing the peace: Experiencing and mitigating emerging harassment in social virtual reality}.
\newblock \bibinfo{journal}{\emph{Proceedings of the ACM on Human-Computer Interaction}} \bibinfo{volume}{6}, \bibinfo{number}{CSCW1} (\bibinfo{year}{2022}), \bibinfo{pages}{1--30}.
\newblock
\urldef\tempurl%
\url{https://doi.org/10.1145/3512932}
\showURL{%
\tempurl}


\bibitem[Furr et~al\mbox{.}(2016)]%
        {furr2016strategic}
\bibfield{author}{\bibinfo{person}{June~B. Furr}, \bibinfo{person}{Alexis Carreiro}, {and} \bibinfo{person}{John~A. McArthur}.} \bibinfo{year}{2016}\natexlab{}.
\newblock \showarticletitle{Strategic Approaches to Disability Disclosure on Social Media}.
\newblock \bibinfo{journal}{\emph{Disability \& Society}} \bibinfo{volume}{31}, \bibinfo{number}{10} (\bibinfo{year}{2016}), \bibinfo{pages}{1353--1368}.
\newblock
\urldef\tempurl%
\url{https://doi.org/10.1080/09687599.2016.1256272}
\showDOI{\tempurl}


\bibitem[Gignac et~al\mbox{.}(2022)]%
        {gignac2022workplace}
\bibfield{author}{\bibinfo{person}{Monique A~M Gignac}, \bibinfo{person}{Julie Bowring}, \bibinfo{person}{Faraz~V Shahidi}, \bibinfo{person}{Vicki Kristman}, \bibinfo{person}{Jill~I Cameron}, {and} \bibinfo{person}{Arif Jetha}.} \bibinfo{year}{2022}\natexlab{}.
\newblock \showarticletitle{{Workplace Disclosure Decisions of Older Workers Wanting to Remain Employed: A Qualitative Study of Factors Considered When Contemplating Revealing or Concealing Support Needs}}.
\newblock \bibinfo{journal}{\emph{Work, Aging and Retirement}} \bibinfo{volume}{10}, \bibinfo{number}{2} (\bibinfo{date}{09} \bibinfo{year}{2022}), \bibinfo{pages}{174--187}.
\newblock
\showISSN{2054-4650}
\urldef\tempurl%
\url{https://doi.org/10.1093/workar/waac029}
\showDOI{\tempurl}
\showeprint{https://academic.oup.com/workar/article-pdf/10/2/174/57119663/waac029.pdf}


\bibitem[Gualano et~al\mbox{.}(2023)]%
        {gualano2023invisible}
\bibfield{author}{\bibinfo{person}{Ria~J Gualano}, \bibinfo{person}{Lucy Jiang}, \bibinfo{person}{Kexin Zhang}, \bibinfo{person}{Andrea~Stevenson Won}, {and} \bibinfo{person}{Shiri Azenkot}.} \bibinfo{year}{2023}\natexlab{}.
\newblock \showarticletitle{“Invisible Illness Is No Longer Invisible”: Making Social VR Avatars More Inclusive for Invisible Disability Representation}. In \bibinfo{booktitle}{\emph{Proceedings of the 25th International ACM SIGACCESS Conference on Computers and Accessibility}}. \bibinfo{pages}{1--4}.
\newblock
\urldef\tempurl%
\url{https://doi.org/10.1145/3597638.3614480}
\showURL{%
\tempurl}


\bibitem[Harley(2022)]%
        {harley2022would}
\bibfield{author}{\bibinfo{person}{Daniel Harley}.} \bibinfo{year}{2022}\natexlab{}.
\newblock \showarticletitle{“This would be sweet in VR”: On the discursive newness of virtual reality}.
\newblock \bibinfo{journal}{\emph{New Media \& Society}} (\bibinfo{year}{2022}), \bibinfo{pages}{14614448221084655}.
\newblock
\urldef\tempurl%
\url{https://doi.org/10.1177/14614448221084655}
\showURL{%
\tempurl}


\bibitem[Harrison et~al\mbox{.}(2002)]%
        {harrison2002role}
\bibfield{author}{\bibinfo{person}{A Harrison}, \bibinfo{person}{G Derwent}, \bibinfo{person}{A Enticknap}, \bibinfo{person}{FD Rose}, {and} \bibinfo{person}{EA Attree}.} \bibinfo{year}{2002}\natexlab{}.
\newblock \showarticletitle{The role of virtual reality technology in the assessment and training of inexperienced powered wheelchair users}.
\newblock \bibinfo{journal}{\emph{Disability and rehabilitation}} \bibinfo{volume}{24}, \bibinfo{number}{11-12} (\bibinfo{year}{2002}), \bibinfo{pages}{599--606}.
\newblock
\urldef\tempurl%
\url{https://doi.org/10.1080/09638280110111360}
\showURL{%
\tempurl}


\bibitem[Heung et~al\mbox{.}(2024)]%
        {heung2024vulnerable}
\bibfield{author}{\bibinfo{person}{Sharon Heung}, \bibinfo{person}{Lucy Jiang}, \bibinfo{person}{Shiri Azenkot}, {and} \bibinfo{person}{Aditya Vashistha}.} \bibinfo{year}{2024}\natexlab{}.
\newblock \showarticletitle{“Vulnerable, Victimized, and Objectified”: Understanding Ableist Hate and Harassment Experienced by Disabled Content Creators on Social Media}.
\newblock  (\bibinfo{year}{2024}).
\newblock
\urldef\tempurl%
\url{https://doi.org/10.1145/3613904.3641949}
\showDOI{\tempurl}


\bibitem[{Hidden Disabilities}(2024)]%
        {sunflower}
\bibfield{author}{\bibinfo{person}{{Hidden Disabilities}}.} \bibinfo{year}{2024}\natexlab{}.
\newblock \showarticletitle{HDS - Global}.
\newblock  (\bibinfo{year}{2024}).
\newblock
\urldef\tempurl%
\url{https://hdsunflower.com/}
\showURL{%
\tempurl}


\bibitem[{Invisible Disabilities Association}(2023)]%
        {invisibledisabilitydef}
\bibfield{author}{\bibinfo{person}{{Invisible Disabilities Association}}.} \bibinfo{year}{2023}\natexlab{}.
\newblock \showarticletitle{What is an Invisible Disability?}
\newblock  (\bibinfo{year}{2023}).
\newblock
\urldef\tempurl%
\url{https://invisibledisabilities.org/what-is-an-invisible-disability/}
\showURL{%
\tempurl}
\newblock
\shownote{Last Accessed: July 6, 2023}.


\bibitem[Jaeger et~al\mbox{.}(2013)]%
        {jaeger2013cyber}
\bibfield{author}{\bibinfo{person}{Lennart Jaeger}, \bibinfo{person}{Julia Kr{\"o}nung}, {and} \bibinfo{person}{Arne Kupetz}.} \bibinfo{year}{2013}\natexlab{}.
\newblock \showarticletitle{Cyber-Me--Analyzing the Effects of Perceived Stigma of Physically Disabled People on the Disguise of the Real Self in Virtual Environments}.
\newblock  (\bibinfo{year}{2013}).
\newblock
\urldef\tempurl%
\url{https://aisel.aisnet.org/icis2013/proceedings/ConferenceTheme/2/}
\showURL{%
\tempurl}


\bibitem[JustSleightly(2024)]%
        {sleighball}
\bibfield{author}{\bibinfo{person}{JustSleightly}.} \bibinfo{year}{2024}\natexlab{}.
\newblock \bibinfo{booktitle}{\emph{SleightlyBall System}}.
\newblock
\urldef\tempurl%
\url{https://github.com/JustSleightly/SleightlyBall}
\showURL{%
\tempurl}


\bibitem[Kao et~al\mbox{.}(2022)]%
        {kao2022audio}
\bibfield{author}{\bibinfo{person}{Dominic Kao}, \bibinfo{person}{Rabindra Ratan}, \bibinfo{person}{Christos Mousas}, \bibinfo{person}{Amogh Joshi}, {and} \bibinfo{person}{Edward~F Melcer}.} \bibinfo{year}{2022}\natexlab{}.
\newblock \showarticletitle{Audio matters too: How audial avatar customization enhances visual avatar customization}. In \bibinfo{booktitle}{\emph{Proceedings of the 2022 CHI Conference on Human Factors in Computing Systems}}. \bibinfo{pages}{1--27}.
\newblock
\urldef\tempurl%
\url{https://doi.org/10.1145/3491102.3501848}
\showURL{%
\tempurl}


\bibitem[Kaur and Saukko(2022)]%
        {kaur2022social}
\bibfield{author}{\bibinfo{person}{Herminder Kaur} {and} \bibinfo{person}{Paula Saukko}.} \bibinfo{year}{2022}\natexlab{}.
\newblock \showarticletitle{Social access: role of digital media in social relations of young people with disabilities}.
\newblock \bibinfo{journal}{\emph{new media \& society}} \bibinfo{volume}{24}, \bibinfo{number}{2} (\bibinfo{year}{2022}), \bibinfo{pages}{420--436}.
\newblock
\urldef\tempurl%
\url{https://doi.org/10.1177/14614448211063177}
\showURL{%
\tempurl}


\bibitem[Kilteni et~al\mbox{.}(2012)]%
        {kilteni2012sense}
\bibfield{author}{\bibinfo{person}{Konstantina Kilteni}, \bibinfo{person}{Raphaela Groten}, {and} \bibinfo{person}{Mel Slater}.} \bibinfo{year}{2012}\natexlab{}.
\newblock \showarticletitle{The Sense of Embodiment in Virtual Reality}.
\newblock \bibinfo{journal}{\emph{Presence}} \bibinfo{volume}{21}, \bibinfo{number}{4} (\bibinfo{year}{2012}), \bibinfo{pages}{373--387}.
\newblock
\urldef\tempurl%
\url{https://doi.org/10.1162/PRES_a_00124}
\showDOI{\tempurl}


\bibitem[Langener et~al\mbox{.}(2022)]%
        {langener2022immersive}
\bibfield{author}{\bibinfo{person}{Simon Langener}, \bibinfo{person}{Randy Klaassen}, \bibinfo{person}{Joanne VanDerNagel}, {and} \bibinfo{person}{Dirk Heylen}.} \bibinfo{year}{2022}\natexlab{}.
\newblock \showarticletitle{Immersive Virtual Reality Avatars for Embodiment Illusions in People With Mild to Borderline Intellectual Disability: User-Centered Development and Feasibility Study}.
\newblock \bibinfo{journal}{\emph{JMIR Serious Games}} \bibinfo{volume}{10}, \bibinfo{number}{4} (\bibinfo{year}{2022}).
\newblock
\urldef\tempurl%
\url{https://doi.org/10.2196/39966}
\showDOI{\tempurl}


\bibitem[Laufer and Weiss(2011)]%
        {laufer2011virtual}
\bibfield{author}{\bibinfo{person}{Yocheved Laufer} {and} \bibinfo{person}{Patrice Tamar~L Weiss}.} \bibinfo{year}{2011}\natexlab{}.
\newblock \showarticletitle{Virtual reality in the assessment and treatment of children with motor impairment: a systematic review}.
\newblock \bibinfo{journal}{\emph{Journal of Physical Therapy Education}} \bibinfo{volume}{25}, \bibinfo{number}{1} (\bibinfo{year}{2011}), \bibinfo{pages}{59--71}.
\newblock


\bibitem[Lee(2014)]%
        {lee2014does}
\bibfield{author}{\bibinfo{person}{Jong-Eun~Roselyn Lee}.} \bibinfo{year}{2014}\natexlab{}.
\newblock \showarticletitle{Does virtual diversity matter?: Effects of avatar-based diversity representation on willingness to express offline racial identity and avatar customization}.
\newblock \bibinfo{journal}{\emph{Computers in Human Behavior}}  \bibinfo{volume}{36} (\bibinfo{year}{2014}), \bibinfo{pages}{190--197}.
\newblock
\showISSN{0747-5632}
\urldef\tempurl%
\url{https://doi.org/10.1016/j.chb.2014.03.040}
\showDOI{\tempurl}


\bibitem[Lee and Park(2011)]%
        {lee2011whose}
\bibfield{author}{\bibinfo{person}{Jong-Eun~Roselyn Lee} {and} \bibinfo{person}{Sung~Gwan Park}.} \bibinfo{year}{2011}\natexlab{}.
\newblock \showarticletitle{"Whose Second Life Is This?" How Avatar-Based Racial Cues Shape Ethno-Racial Minorities' Perception of Virtual Worlds}.
\newblock \bibinfo{journal}{\emph{Cyberpsychology, Behavior, and Social Networking}} \bibinfo{volume}{14}, \bibinfo{number}{11} (\bibinfo{year}{2011}), \bibinfo{pages}{677--683}.
\newblock
\urldef\tempurl%
\url{https://doi.org/10.1089/cyber.2010.0501}
\showDOI{\tempurl}


\bibitem[Lindsay et~al\mbox{.}(2018)]%
        {lindsay2018systematic}
\bibfield{author}{\bibinfo{person}{Sally Lindsay}, \bibinfo{person}{Elaine Cagliostro}, {and} \bibinfo{person}{Gabriella Carafa}.} \bibinfo{year}{2018}\natexlab{}.
\newblock \showarticletitle{A Systematic Review of Barriers and Facilitators of Disability Disclosure and Accommodations for Youth in Post-Secondary Education}.
\newblock \bibinfo{journal}{\emph{International Journal of Disability, Development and Education}} \bibinfo{volume}{65}, \bibinfo{number}{5} (\bibinfo{year}{2018}), \bibinfo{pages}{526--556}.
\newblock
\urldef\tempurl%
\url{https://doi.org/10.1080/1034912X.2018.1430352}
\showDOI{\tempurl}


\bibitem[Lindsay et~al\mbox{.}(2019)]%
        {lindsay2019disability}
\bibfield{author}{\bibinfo{person}{Sally Lindsay}, \bibinfo{person}{Elaine Cagliostro}, \bibinfo{person}{Joanne Leck}, \bibinfo{person}{Winny Shen}, {and} \bibinfo{person}{Jennifer Stinson}.} \bibinfo{year}{2019}\natexlab{}.
\newblock \showarticletitle{Disability Disclosure and Workplace Accommodations Among Youth With Disabilities}.
\newblock \bibinfo{journal}{\emph{Disability and Rehabilitation}} \bibinfo{volume}{41}, \bibinfo{number}{16} (\bibinfo{year}{2019}), \bibinfo{pages}{1914--1924}.
\newblock
\urldef\tempurl%
\url{https://doi.org/10.1080/09638288.2018.1451926}
\showDOI{\tempurl}
\newblock
\shownote{PMID: 29558221}.


\bibitem[Mack et~al\mbox{.}(2023)]%
        {mack2023towards}
\bibfield{author}{\bibinfo{person}{Kelly~Avery Mack}, \bibinfo{person}{Rai Ching~Ling Hsu}, \bibinfo{person}{Andr\'{e}s Monroy-Hern\'{a}ndez}, \bibinfo{person}{Brian~A. Smith}, {and} \bibinfo{person}{Fannie Liu}.} \bibinfo{year}{2023}\natexlab{}.
\newblock \showarticletitle{Towards Inclusive Avatars: Disability Representation in Avatar Platforms}. In \bibinfo{booktitle}{\emph{Proceedings of the 2023 CHI Conference on Human Factors in Computing Systems}} (Hamburg, Germany) \emph{(\bibinfo{series}{CHI '23})}. \bibinfo{publisher}{Association for Computing Machinery}, \bibinfo{address}{New York, NY, USA}, Article \bibinfo{articleno}{607}, \bibinfo{numpages}{13}~pages.
\newblock
\showISBNx{9781450394215}
\urldef\tempurl%
\url{https://doi.org/10.1145/3544548.3581481}
\showDOI{\tempurl}


\bibitem[Maloney and Freeman(2020)]%
        {maloney2020falling}
\bibfield{author}{\bibinfo{person}{Divine Maloney} {and} \bibinfo{person}{Guo Freeman}.} \bibinfo{year}{2020}\natexlab{}.
\newblock \showarticletitle{Falling Asleep Together: What Makes Activities in Social Virtual Reality Meaningful to Users}. In \bibinfo{booktitle}{\emph{Proceedings of the Annual Symposium on Computer-Human Interaction in Play}} (Virtual Event, Canada) \emph{(\bibinfo{series}{CHI PLAY '20})}. \bibinfo{publisher}{Association for Computing Machinery}, \bibinfo{address}{New York, NY, USA}, \bibinfo{pages}{510–521}.
\newblock
\showISBNx{9781450380744}
\urldef\tempurl%
\url{https://doi.org/10.1145/3410404.3414266}
\showDOI{\tempurl}


\bibitem[Mason et~al\mbox{.}(2022)]%
        {mason2022including}
\bibfield{author}{\bibinfo{person}{Liam Mason}, \bibinfo{person}{Kathrin Gerling}, \bibinfo{person}{Patrick Dickinson}, \bibinfo{person}{Jussi Holopainen}, \bibinfo{person}{Lisa Jacobs}, {and} \bibinfo{person}{Kieran Hicks}.} \bibinfo{year}{2022}\natexlab{}.
\newblock \showarticletitle{Including the experiences of physically disabled players in mainstream guidelines for movement-based games}. In \bibinfo{booktitle}{\emph{Proceedings of the 2022 CHI Conference on Human Factors in Computing Systems}}. \bibinfo{pages}{1--15}.
\newblock
\urldef\tempurl%
\url{https://doi.org/10.1145/3491102.3501867}
\showURL{%
\tempurl}


\bibitem[Matamala-Gomez et~al\mbox{.}(2019)]%
        {matamala2019immersive}
\bibfield{author}{\bibinfo{person}{Marta Matamala-Gomez}, \bibinfo{person}{Tony Donegan}, \bibinfo{person}{Sara Bottiroli}, \bibinfo{person}{Giorgio Sandrini}, \bibinfo{person}{Maria~V Sanchez-Vives}, {and} \bibinfo{person}{Cristina Tassorelli}.} \bibinfo{year}{2019}\natexlab{}.
\newblock \showarticletitle{Immersive virtual reality and virtual embodiment for pain relief}.
\newblock \bibinfo{journal}{\emph{Frontiers in human neuroscience}}  \bibinfo{volume}{13} (\bibinfo{year}{2019}), \bibinfo{pages}{279}.
\newblock
\urldef\tempurl%
\url{https://doi.org/10.3389/fnhum.2019.00279}
\showURL{%
\tempurl}


\bibitem[McGrath et~al\mbox{.}(2023)]%
        {mcgrath2023disclosure}
\bibfield{author}{\bibinfo{person}{Martina~O McGrath}, \bibinfo{person}{Karolina Krysinska}, \bibinfo{person}{Nicola~J Reavley}, \bibinfo{person}{Karl Andriessen}, {and} \bibinfo{person}{Jane Pirkis}.} \bibinfo{year}{2023}\natexlab{}.
\newblock \showarticletitle{Disclosure of mental health problems or suicidality at work: a systematic review}.
\newblock \bibinfo{journal}{\emph{International journal of environmental research and public health}} \bibinfo{volume}{20}, \bibinfo{number}{8} (\bibinfo{year}{2023}), \bibinfo{pages}{5548}.
\newblock
\urldef\tempurl%
\url{https://doi.org/10.3390/ijerph20085548}
\showURL{%
\tempurl}


\bibitem[Miserandino(2017)]%
        {spoontheory}
\bibfield{author}{\bibinfo{person}{Christine Miserandino}.} \bibinfo{year}{2017}\natexlab{}.
\newblock \bibinfo{booktitle}{\emph{The Spoon Theory}}.
\newblock \bibinfo{publisher}{Routledge}. 174--178 pages.
\newblock


\bibitem[Moriña(2022)]%
        {morina2022when}
\bibfield{author}{\bibinfo{person}{Anabel Moriña}.} \bibinfo{year}{2022}\natexlab{}.
\newblock \showarticletitle{When What is Unseen Does Not Exist: Disclosure, Barriers and Supports for Students With Invisible Disabilities in Higher Education}.
\newblock \bibinfo{journal}{\emph{Disability \& Society}} (\bibinfo{year}{2022}), \bibinfo{pages}{1--19}.
\newblock
\urldef\tempurl%
\url{https://doi.org/10.1080/09687599.2022.2113038}
\showDOI{\tempurl}


\bibitem[Morris et~al\mbox{.}(2023)]%
        {morris2023don}
\bibfield{author}{\bibinfo{person}{Margaret~E Morris}, \bibinfo{person}{Daniela~K Rosner}, \bibinfo{person}{Paula~S Nurius}, {and} \bibinfo{person}{Hadar~M Dolev}.} \bibinfo{year}{2023}\natexlab{}.
\newblock \showarticletitle{“I Don't Want to Hide Behind an Avatar”: Self-Representation in Social VR Among Women in Midlife}. In \bibinfo{booktitle}{\emph{Proceedings of the 2023 ACM Designing Interactive Systems Conference}} (Pittsburgh, PA, USA) \emph{(\bibinfo{series}{DIS '23})}. \bibinfo{publisher}{Association for Computing Machinery}, \bibinfo{address}{New York, NY, USA}, \bibinfo{pages}{537–546}.
\newblock
\showISBNx{9781450398930}
\urldef\tempurl%
\url{https://doi.org/10.1145/3563657.3596129}
\showDOI{\tempurl}


\bibitem[Motley(2021)]%
        {youtube_invisibility_toolbox}
\bibfield{author}{\bibinfo{person}{Toolbox Motley}.} \bibinfo{year}{2021}\natexlab{}.
\newblock \bibinfo{title}{How to add invisibility to your VRChat avatar.}
\newblock \bibinfo{howpublished}{\url{https://www.youtube.com/watch?v=MlmwPnNBt_g}}.
\newblock
\newblock
\shownote{Last Accessed: September 14, 2023}.


\bibitem[Mullins and Preyde(2013)]%
        {mullins2013lived}
\bibfield{author}{\bibinfo{person}{Laura Mullins} {and} \bibinfo{person}{Michèle Preyde}.} \bibinfo{year}{2013}\natexlab{}.
\newblock \showarticletitle{The Lived Experience of Students With an Invisible Disability at a Canadian University}.
\newblock \bibinfo{journal}{\emph{Disability \& Society}} \bibinfo{volume}{28}, \bibinfo{number}{2} (\bibinfo{year}{2013}), \bibinfo{pages}{147--160}.
\newblock
\urldef\tempurl%
\url{https://doi.org/10.1080/09687599.2012.752127}
\showDOI{\tempurl}


\bibitem[Pearson et~al\mbox{.}(2003)]%
        {pearson2003to}
\bibfield{author}{\bibinfo{person}{Veronica Pearson}, \bibinfo{person}{F. Ip}, \bibinfo{person}{H. Hui}, \bibinfo{person}{Nelson Yip}, \bibinfo{person}{K.K. Ho}, {and} \bibinfo{person}{E. Lo}.} \bibinfo{year}{2003}\natexlab{}.
\newblock \showarticletitle{To Tell or Not to Tell; Disability Disclosure and Job Application Outcomes}.
\newblock \bibinfo{journal}{\emph{Journal of Rehabilitation}}  \bibinfo{volume}{69} (\bibinfo{date}{10} \bibinfo{year}{2003}), \bibinfo{pages}{35--38}.
\newblock


\bibitem[Porter et~al\mbox{.}(2017)]%
        {porter2017filtered}
\bibfield{author}{\bibinfo{person}{John~R. Porter}, \bibinfo{person}{Kiley Sobel}, \bibinfo{person}{Sarah~E. Fox}, \bibinfo{person}{Cynthia~L. Bennett}, {and} \bibinfo{person}{Julie~A. Kientz}.} \bibinfo{year}{2017}\natexlab{}.
\newblock \showarticletitle{Filtered Out: Disability Disclosure Practices in Online Dating Communities}.
\newblock \bibinfo{journal}{\emph{Proceedings of the ACM on Human-Computer Interaction}} \bibinfo{volume}{1}, \bibinfo{number}{CSCW}, Article \bibinfo{articleno}{87} (\bibinfo{year}{2017}), \bibinfo{numpages}{13}~pages.
\newblock
\urldef\tempurl%
\url{https://doi.org/10.1145/3134722}
\showDOI{\tempurl}


\bibitem[Povinelli and Zhao(2024)]%
        {povinelli2024springboard}
\bibfield{author}{\bibinfo{person}{Kassie Povinelli} {and} \bibinfo{person}{Yuhang Zhao}.} \bibinfo{year}{2024}\natexlab{}.
\newblock \showarticletitle{Springboard, Roadblock or" Crutch"?: How Transgender Users Leverage Voice Changers for Gender Presentation in Social Virtual Reality}.
\newblock \bibinfo{journal}{\emph{arXiv preprint arXiv:2402.08217}} (\bibinfo{year}{2024}).
\newblock
\urldef\tempurl%
\url{https://arxiv.org/abs/2402.08217}
\showURL{%
\tempurl}


\bibitem[{Ready Player Me}(2023)]%
        {readyplayerme}
\bibfield{author}{\bibinfo{person}{{Ready Player Me}}.} \bibinfo{year}{2023}\natexlab{}.
\newblock  (\bibinfo{year}{2023}).
\newblock
\urldef\tempurl%
\url{https://readyplayer.me/}
\showURL{%
\tempurl}
\newblock
\shownote{Accessed: July 6, 2023}.


\bibitem[Ribeiro et~al\mbox{.}(2024)]%
        {ribeiro2024towards}
\bibfield{author}{\bibinfo{person}{Ailton Ribeiro}, \bibinfo{person}{Murilo~Guerreiro Arouca}, \bibinfo{person}{Ana~Maria Amorim}, \bibinfo{person}{Maria~Clara Pestana}, {and} \bibinfo{person}{Vaninha Vieira}.} \bibinfo{year}{2024}\natexlab{}.
\newblock \showarticletitle{Towards inclusive avatars: A study on self-representation in virtual environments}. In \bibinfo{booktitle}{\emph{Anais do XIX Simp{\'o}sio Brasileiro de Sistemas Colaborativos}}. SBC, \bibinfo{pages}{13--27}.
\newblock
\urldef\tempurl%
\url{https://doi.org/10.5753/sbsc.2024.238056}
\showURL{%
\tempurl}


\bibitem[Rocco(2004)]%
        {rocco2004towards}
\bibfield{author}{\bibinfo{person}{Tonette Rocco}.} \bibinfo{year}{2004}\natexlab{}.
\newblock \showarticletitle{Towards A Model Of Disability Disclosure}.
\newblock  (\bibinfo{year}{2004}).
\newblock
\urldef\tempurl%
\url{https://scholarworks.iupui.edu/server/api/core/bitstreams/5b732c98-39ff-4b77-929a-4a1f609bd082/content}
\showURL{%
\tempurl}


\bibitem[Roth et~al\mbox{.}(2016)]%
        {roth2016avatar}
\bibfield{author}{\bibinfo{person}{Daniel Roth}, \bibinfo{person}{Jean-Luc Lugrin}, \bibinfo{person}{Dmitri Galakhov}, \bibinfo{person}{Arvid Hofmann}, \bibinfo{person}{Gary Bente}, \bibinfo{person}{Marc~Erich Latoschik}, {and} \bibinfo{person}{Arnulph Fuhrmann}.} \bibinfo{year}{2016}\natexlab{}.
\newblock \showarticletitle{Avatar realism and social interaction quality in virtual reality}. In \bibinfo{booktitle}{\emph{2016 IEEE virtual reality (VR)}}. IEEE, \bibinfo{pages}{277--278}.
\newblock
\urldef\tempurl%
\url{https://doi.org/10.1109/VR.2016.7504761}
\showURL{%
\tempurl}


\bibitem[Sadeh-Sharvit et~al\mbox{.}(2021)]%
        {sharvit2021virtual}
\bibfield{author}{\bibinfo{person}{Shiri Sadeh-Sharvit}, \bibinfo{person}{Jonathan Giron}, \bibinfo{person}{Shir Fridman}, \bibinfo{person}{Maxine Hanrieder}, \bibinfo{person}{Shany Goldstein}, \bibinfo{person}{Doron Friedman}, {and} \bibinfo{person}{Shir Brokman}.} \bibinfo{year}{2021}\natexlab{}.
\newblock \showarticletitle{Virtual Reality in Sexual Harassment Prevention: Proof-of-Concept Study}. In \bibinfo{booktitle}{\emph{Proceedings of the 21st ACM International Conference on Intelligent Virtual Agents}} (Virtual Event, Japan) \emph{(\bibinfo{series}{IVA '21})}. \bibinfo{publisher}{Association for Computing Machinery}, \bibinfo{address}{New York, NY, USA}, \bibinfo{pages}{87–89}.
\newblock
\showISBNx{9781450386197}
\urldef\tempurl%
\url{https://doi.org/10.1145/3472306.3478356}
\showDOI{\tempurl}


\bibitem[Saltes(2013)]%
        {saltes2013disability}
\bibfield{author}{\bibinfo{person}{Natasha Saltes}.} \bibinfo{year}{2013}\natexlab{}.
\newblock \showarticletitle{Disability, identity and disclosure in the online dating environment}.
\newblock \bibinfo{journal}{\emph{Disability \& Society}} \bibinfo{volume}{28}, \bibinfo{number}{1} (\bibinfo{year}{2013}), \bibinfo{pages}{96--109}.
\newblock
\urldef\tempurl%
\url{https://doi.org/10.1080/09687599.2012.695577}
\showURL{%
\tempurl}


\bibitem[Steed and Schroeder(2015)]%
        {steed2015collaboration}
\bibfield{author}{\bibinfo{person}{Anthony Steed} {and} \bibinfo{person}{Ralph Schroeder}.} \bibinfo{year}{2015}\natexlab{}.
\newblock \showarticletitle{Collaboration in immersive and non-immersive virtual environments}.
\newblock \bibinfo{journal}{\emph{Immersed in media: Telepresence theory, measurement \& technology}} (\bibinfo{year}{2015}), \bibinfo{pages}{263--282}.
\newblock
\urldef\tempurl%
\url{https://doi.org/10.1007/978-3-319-10190-3_11}
\showURL{%
\tempurl}


\bibitem[Tham et~al\mbox{.}(2018)]%
        {tham2018understanding}
\bibfield{author}{\bibinfo{person}{Jason Tham}, \bibinfo{person}{Ann~Hill Duin}, \bibinfo{person}{Laura Gee}, \bibinfo{person}{Nathan Ernst}, \bibinfo{person}{Bilal Abdelqader}, {and} \bibinfo{person}{Megan McGrath}.} \bibinfo{year}{2018}\natexlab{}.
\newblock \showarticletitle{Understanding virtual reality: Presence, embodiment, and professional practice}.
\newblock \bibinfo{journal}{\emph{IEEE Transactions on Professional Communication}} \bibinfo{volume}{61}, \bibinfo{number}{2} (\bibinfo{year}{2018}), \bibinfo{pages}{178--195}.
\newblock
\urldef\tempurl%
\url{https://doi.org/10.1109/TPC.2018.2804238}
\showURL{%
\tempurl}


\bibitem[Toth and Dewa(2014)]%
        {toth2014employee}
\bibfield{author}{\bibinfo{person}{Kate~E Toth} {and} \bibinfo{person}{Carolyn~S Dewa}.} \bibinfo{year}{2014}\natexlab{}.
\newblock \showarticletitle{Employee decision-making about disclosure of a mental disorder at work}.
\newblock \bibinfo{journal}{\emph{Journal of occupational rehabilitation}}  \bibinfo{volume}{24} (\bibinfo{year}{2014}), \bibinfo{pages}{732--746}.
\newblock
\urldef\tempurl%
\url{https://doi.org/10.1007/s10926-014-9504-y}
\showURL{%
\tempurl}


\bibitem[Toth et~al\mbox{.}(2022)]%
        {toth2022disclosure}
\bibfield{author}{\bibinfo{person}{Kate~E Toth}, \bibinfo{person}{Florence Yvon}, \bibinfo{person}{Patrizia Villotti}, \bibinfo{person}{Tania Lecomte}, \bibinfo{person}{Jean-Philippe Lachance}, \bibinfo{person}{Bonnie Kirsh}, \bibinfo{person}{Heather Stuart}, \bibinfo{person}{Djamal Berbiche}, {and} \bibinfo{person}{Marc Corbi{\`e}re}.} \bibinfo{year}{2022}\natexlab{}.
\newblock \showarticletitle{Disclosure dilemmas: how people with a mental health condition perceive and manage disclosure at work}.
\newblock \bibinfo{journal}{\emph{Disability and Rehabilitation}} \bibinfo{volume}{44}, \bibinfo{number}{25} (\bibinfo{year}{2022}), \bibinfo{pages}{7791--7801}.
\newblock
\urldef\tempurl%
\url{https://doi.org/10.1080/09638288.2021.1998667}
\showURL{%
\tempurl}


\bibitem[von Schrader et~al\mbox{.}(2014)]%
        {von2014perspectives}
\bibfield{author}{\bibinfo{person}{Sarah von Schrader}, \bibinfo{person}{Valerie Malzer}, {and} \bibinfo{person}{Susanne Bruyère}.} \bibinfo{year}{2014}\natexlab{}.
\newblock \showarticletitle{Perspectives on Disability Disclosure: The Importance of Employer Practices and Workplace Climate}.
\newblock \bibinfo{journal}{\emph{Employee Responsibilities and Rights Journal}} \bibinfo{volume}{26}, \bibinfo{number}{4} (\bibinfo{year}{2014}), \bibinfo{pages}{237--255}.
\newblock
\showISBNx{1573-3378}
\urldef\tempurl%
\url{https://doi.org/10.1007/s10672-013-9227-9}
\showDOI{\tempurl}


\bibitem[Wilton(2006)]%
        {wilton2006disability}
\bibfield{author}{\bibinfo{person}{Robert~D Wilton}.} \bibinfo{year}{2006}\natexlab{}.
\newblock \showarticletitle{Disability Disclosure in the Workplace}.
\newblock \bibinfo{journal}{\emph{Just Labour}}  \bibinfo{volume}{8} (\bibinfo{year}{2006}).
\newblock
\urldef\tempurl%
\url{https://doi.org/10.25071/1705-1436.107}
\showDOI{\tempurl}


\bibitem[Won and Davis(2023)]%
        {won2023your}
\bibfield{author}{\bibinfo{person}{Andrea~Stevenson Won} {and} \bibinfo{person}{Donna~Z Davis}.} \bibinfo{year}{2023}\natexlab{}.
\newblock \showarticletitle{Your money or your data: Avatar embodiment options in the identity economy}.
\newblock \bibinfo{journal}{\emph{Convergence}} (\bibinfo{year}{2023}), \bibinfo{pages}{13548565231200187}.
\newblock


\bibitem[World(2023)]%
        {disabledworld}
\bibfield{author}{\bibinfo{person}{Disabled World}.} \bibinfo{year}{2023}\natexlab{}.
\newblock \showarticletitle{Invisible Disabilities: List and General Information}.
\newblock  (\bibinfo{year}{2023}).
\newblock
\urldef\tempurl%
\url{https://www.disabled-world.com/disability/types/invisible/}
\showURL{%
\tempurl}
\newblock
\shownote{Last Accessed: July 6, 2023}.


\bibitem[Yalon-Chamovitz and Weiss(2008)]%
        {yalon2008virtual}
\bibfield{author}{\bibinfo{person}{Shira Yalon-Chamovitz} {and} \bibinfo{person}{Patrice L.~(Tamar) Weiss}.} \bibinfo{year}{2008}\natexlab{}.
\newblock \showarticletitle{Virtual Reality as a Leisure Activity for Young Adults With Physical and Intellectual Disabilities}.
\newblock \bibinfo{journal}{\emph{Research in Developmental Disabilities}} \bibinfo{volume}{29}, \bibinfo{number}{3} (\bibinfo{year}{2008}), \bibinfo{pages}{273--287}.
\newblock
\showISSN{0891-4222}
\urldef\tempurl%
\url{https://doi.org/10.1016/j.ridd.2007.05.004}
\showDOI{\tempurl}


\bibitem[Zhang et~al\mbox{.}(2022)]%
        {zhang2022s}
\bibfield{author}{\bibinfo{person}{Kexin Zhang}, \bibinfo{person}{Elmira Deldari}, \bibinfo{person}{Zhicong Lu}, \bibinfo{person}{Yaxing Yao}, {and} \bibinfo{person}{Yuhang Zhao}.} \bibinfo{year}{2022}\natexlab{}.
\newblock \showarticletitle{“It’s Just Part of Me:” Understanding Avatar Diversity and Self-Presentation of People with Disabilities in Social Virtual Reality}. In \bibinfo{booktitle}{\emph{Proceedings of the 24th International ACM SIGACCESS Conference on Computers and Accessibility}} (Athens, Greece) \emph{(\bibinfo{series}{ASSETS '22})}. \bibinfo{publisher}{Association for Computing Machinery}, \bibinfo{address}{New York, NY, USA}, Article \bibinfo{articleno}{4}, \bibinfo{numpages}{16}~pages.
\newblock
\showISBNx{9781450392587}
\urldef\tempurl%
\url{https://doi.org/10.1145/3517428.3544829}
\showDOI{\tempurl}


\bibitem[Zhang et~al\mbox{.}(2023)]%
        {zhang2023diary}
\bibfield{author}{\bibinfo{person}{Kexin Zhang}, \bibinfo{person}{Elmira Deldari}, \bibinfo{person}{Yaxing Yao}, {and} \bibinfo{person}{Yuhang Zhao}.} \bibinfo{year}{2023}\natexlab{}.
\newblock \showarticletitle{A Diary Study in Social Virtual Reality: Impact of Avatars with Disability Signifiers on the Social Experiences of People with Disabilities}. In \bibinfo{booktitle}{\emph{Proceedings of the 25th International ACM SIGACCESS Conference on Computers and Accessibility}} (, New York, NY, USA,) \emph{(\bibinfo{series}{ASSETS '23})}. \bibinfo{publisher}{Association for Computing Machinery}, \bibinfo{address}{New York, NY, USA}, Article \bibinfo{articleno}{40}, \bibinfo{numpages}{17}~pages.
\newblock
\showISBNx{9798400702204}
\urldef\tempurl%
\url{https://doi.org/10.1145/3597638.3608388}
\showDOI{\tempurl}


\bibitem[Zhang et~al\mbox{.}(2021)]%
        {zhang2021social}
\bibfield{author}{\bibinfo{person}{Lotus Zhang}, \bibinfo{person}{Lucy Jiang}, \bibinfo{person}{Nicole Washington}, \bibinfo{person}{Augustina~Ao Liu}, \bibinfo{person}{Jingyao Shao}, \bibinfo{person}{Adam Fourney}, \bibinfo{person}{Meredith~Ringel Morris}, {and} \bibinfo{person}{Leah Findlater}.} \bibinfo{year}{2021}\natexlab{}.
\newblock \showarticletitle{Social media through voice: Synthesized voice qualities and self-presentation}.
\newblock \bibinfo{journal}{\emph{Proceedings of the ACM on Human-Computer Interaction}} \bibinfo{volume}{5}, \bibinfo{number}{CSCW1} (\bibinfo{year}{2021}), \bibinfo{pages}{1--21}.
\newblock
\urldef\tempurl%
\url{https://doi.org/10.1145/3449235}
\showURL{%
\tempurl}


\end{thebibliography}

\end{document}